\documentclass[twocolumn]{aastex631}

\usepackage{epsfig}
\usepackage{natbib}
\usepackage{amsmath}
\usepackage{hyperref}
\usepackage{comment}
\usepackage{multirow,tabularx}
\usepackage[flushleft]{threeparttable}
\usepackage{amssymb}
\usepackage{soul}

\usepackage{xcolor}
\defcitealias{Jones_2017_I}{J17}
\defcitealias{hlozek12}{H12}
\defcitealias{Shivvers_2017}{S17}
\defcitealias{kessler19_DESbiascor}{K19}
\defcitealias{Vincenzi_2019}{V19} % Not used
\defcitealias{vincenzi21}{V21}
\defcitealias{popovic21_dust2dust}{P21-dust}
\defcitealias{brout21_BS20}{BS20}
\defcitealias{DES-CC}{DES-nonIa}

\def\snana{\texttt{SNANA}}

\reportnum{DES-2023-805}
\reportnum{FERMILAB-PUB-23-0821-PPD}

\def\numdeshd{1635} %number on hubble diagram (ignoring classification)

\def\numdesia{1499} %number on hubble diagram with pia>0.5
\def\numhubble{1829} %number on hubble diagram 

\def\numlowzia{194} %number on hubble diagram 

\def\FlcdmPLANCKomegam{$0.317\pm0.008$} 
 
\def\mflcdmSN{$0.352\pm 0.017$}
\def\mflcdmSNP{$0.338^{+0.016}_{-0.014}$}
\def\mflcdmSNBpt{$0.330^{+0.011}_{-0.010}$}
\def\mflcdmSNPBpt{$0.315\pm0.007$}

\def\mlcdmSN{$0.291^{+0.063}_{-0.065}$}
\def\mlcdmSNP{$0.359^{+0.014}_{-0.016}$}
\def\mlcdmSNBpt{$0.327^{+0.012}_{-0.011}$}
\def\mlcdmSNPBpt{$0.318^{+0.011}_{-0.010}$}

\def\klcdmSN{$0.16\pm 0.16$}
\def\klcdmSNP{${-}0.010\pm 0.005$}
\def\klcdmSNBpt{$0.030\pm 0.034$}
\def\klcdmSNPBpt{$0.002^{+0.004}_{-0.003}$}

\def\llcdmSN{$0.55\pm 0.10$} %\pm0.17$} % Original text had an overestimate in the uncertainty, having ignored correlations between the parameters. Taking into account correlations means the uncertainty is much smaller than when they are ignored! Tamara's mistake while drafting the paragraph (we fit for Omega_K rather than Omega_Lambda), pointed out by Ósmar Rodríguez. This only appears in 4.1.2 text, not in any tables. 
\def\qFlatcosmographic{$-0.530^{+0.018}_{-0.017}$}

\def\mfwcdmSN{$0.264^{+0.074}_{-0.096}$}
\def\mfwcdmSNP{$0.337^{+0.013}_{-0.011}$}
\def\mfwcdmSNBpt{$0.323^{+0.011}_{-0.010}$}
\def\mfwcdmSNPBpt{$0.321\pm0.007$}

\def\wfwcdmSN{$-0.80^{+0.14}_{-0.16}$}
\def\wfwcdmSNP{$-0.955^{+0.032}_{-0.037}$}
\def\wfwcdmSNBpt{$-0.922^{+0.035}_{-0.037}$}
\def\wfwcdmSNPBpt{$-0.941\pm0.026$}

\def\mfwaSN{$0.495^{+0.033}_{-0.043}$}
\def\mfwaSNP{$0.325^{+0.016}_{-0.012}$}
\def\mfwaSNBpt{$0.334\pm 0.012$}
\def\mfwaSNPBpt{$0.325\pm 0.008$}

\def\wfwaSN{$-0.36^{+0.36}_{-0.30}$}
\def\wfwaSNP{$-0.73\pm 0.11$}
\def\wfwaSNBpt{$-0.778^{+0.088}_{-0.080}$}
\def\wfwaSNPBpt{$-0.773^{+0.075}_{-0.067}$}

\def\wafwaSN{$-8.8^{+3.7}_{-4.5}$}
\def\wafwaSNP{$-1.17^{+0.55}_{-0.62}$}
\def\wafwaSNBpt{$-0.93^{+0.46}_{-0.53}$}
\def\wafwaSNPBpt{$-0.83^{+0.33}_{-0.42}$}

\def\FLCDMomNOLOWZ{$0.330\pm0.024$}
\def\FLCDMomNOHZ{$0.342\pm0.017$}
\def\FLCDMDomNOLOWZ{$-0.022$}
\def\FLCDMDomNOHZ{$-0.010$}

\def\FLCDMsigNOLOWZ{$1.1$}
\def\FLCDMsigNOHZ{$2.1$}

\def\wCDMomNOLOWZ{$0.373\pm0.058$}
\def\wCDMomNOHZ{$0.139\pm0.088$}
\def\wCDMwNOLOWZ{$-1.34\pm0.32$}
\def\wCDMwNOHZ{$-0.66\pm0.11$}
\def\wCDMDwNOLOWZ{$0.54$}
\def\wCDMDwNOHZ{$-0.14$}
\def\CMBwCDMomNOLOWZ{$0.321\pm0.013$}
\def\CMBwCDMomNOHZ{$0.328\pm0.010$}
\def\CMBwCDMwNOLOWZ{$-0.985\pm0.048$}
\def\CMBwCDMwNOHZ{$-0.951\pm0.030$}
\def\CMBwCDMDwNOLOWZ{$0.043$}
\def\CMBwCDMDwNOHZ{$0.009$}

\def\wCDMsigNOLOWZ{$2.3$}
\def\wCDMsigNOHZ{$-2.2$}
\def\CMBwCDMsigNOLOWZ{$1.1$}
\def\CMBwCDMsigNOHZ{$1.9$}

\def\waCDMomNOLOWZ{$0.460\pm0.100$}
\def\waCDMomNOHZ{$0.363\pm0.123$}
\def\waCDMwNOLOWZ{$-0.58\pm0.74$}
\def\waCDMwNOHZ{$-0.58\pm0.18$}
\def\waCDMwaNOLOWZ{$-6.9\pm6.0$}
\def\waCDMwaNOHZ{$-3.7\pm3.2$}
\def\waCDMDwNOLOWZ{$0.22$}
\def\waCDMDwNOHZ{$0.22$}
\def\waCDMDwaNOLOWZ{$-1.9$}
\def\waCDMDwaNOHZ{$-5.1$}

\newcommand{\BAOpt}{BAO+3$\times$2pt}

\def\DeltaBE{$\Delta$(ln $BE$)}

\newcommand{\zr}{\bar{z}}
\newcommand{\om}{\Omega_{\rm M}}
\newcommand{\oll}{\Omega_{\rm \Lambda}}
\newcommand{\ok}{\Omega_{\rm K}}

\newcommand{\mate}{Q_H}

\newcommand{\mubias}{\Delta\mu_{{\rm bias},i}}
\newcommand{\kmsMpc}{km\,s$^{-1}$Mpc$^{-1}$}
\newcommand{\beq}{\begin{equation}}
\newcommand{\eeq}{\end{equation}}
\newcommand{\bea}{\begin{eqnarray}}
\newcommand{\eea}{\end{eqnarray}}
\newcommand{\dl}{D_{\rm L}}

\newcommand{\PBIa}{P_{{\rm B(Ia)}}}

\shorttitle{The Dark Energy Survey: supernova cosmology results}
\shortauthors{The Dark Energy Survey Collaboration}
\begin{document}

\title{
The Dark Energy Survey: Cosmology Results With $\sim$1500 New High-redshift Type Ia Supernovae Using The Full 5-year Dataset}

\author{DES Collaboration: T.~M.~C.~Abbott}
\affiliation{Cerro Tololo Inter-American Observatory, NSF's National Optical-Infrared Astronomy Research Laboratory, Casilla 603, La Serena, Chile}
\author{M.~Acevedo}
\affiliation{Department of Physics, Duke University Durham, NC 27708, USA}
\author{M.~Aguena}
\affiliation{Laborat\'orio Interinstitucional de e-Astronomia - LIneA, Rua Gal. Jos\'e Cristino 77, Rio de Janeiro, RJ - 20921-400, Brazil}
\author{A.~Alarcon}
\affiliation{Argonne National Laboratory, 9700 South Cass Avenue, Lemont, IL 60439, USA}
\author{S.~Allam}
\affiliation{Fermi National Accelerator Laboratory, P. O. Box 500, Batavia, IL 60510, USA}
\author{O.~Alves}
\affiliation{Department of Physics, University of Michigan, Ann Arbor, MI 48109, USA}
\author{A.~Amon}
\affiliation{Department of Astrophysical Sciences, Princeton University, Princeton, NJ 08544}
\author{F.~Andrade-Oliveira}
\affiliation{Department of Physics, University of Michigan, Ann Arbor, MI 48109, USA}
\author{J.~Annis}
\affiliation{Fermi National Accelerator Laboratory, P. O. Box 500, Batavia, IL 60510, USA}
\author{P.~Armstrong}
\affiliation{The Research School of Astronomy and Astrophysics, Australian National University, ACT 2601, Australia}
% ...
%\collaboration{{(DES Collaboration)}}

%\author[0000-0002-0786-7307]{The Dark Energy Survey}
%\affiliation{Many Institutions}

%% Second arxiv submission added authors E. Kasai and U. Malik.

%\author{Alphabetical Order}
%\affiliation{First affiliation}

%\author{Alpha Order}
%\affiliation{Second affiliation}
%\affiliation{Third affiliation}

%\author{Alpha Ordering}
%\altaffiliation{Extra affiliation details}
%\affiliation{Fourth affiliation}

%\collaboration{20}{(The Dark Energy Survey)}



\begin{abstract}
%{Intended for ApJL.  250 word limit}
We present cosmological constraints from the sample of Type Ia supernovae (SN~Ia) discovered and measured during the full five years of the Dark Energy Survey (DES) Supernova Program. In contrast to most previous cosmological samples, in which supernovae are classified based on their spectra, we classify the DES supernovae using a machine learning algorithm applied to their light curves in four photometric bands. Spectroscopic redshifts are acquired from a dedicated follow-up survey of the host galaxies. After accounting for the likelihood of each supernova being a SN~Ia, we find 1635 DES~SNe in the redshift-range $0.10<z<1.13$ that pass quality selection criteria sufficient to constrain cosmological parameters.
This quintuples the number of high-quality $z>0.5$ SNe compared to the previous leading compilation of Pantheon+, and results in the tightest cosmological constraints achieved by any supernova data set to date. To derive cosmological constraints we combine the DES supernova data with a high-quality external low-redshift sample consisting of 194~SNe~Ia spanning $0.025<z<0.10$.
Using supernova data alone and including systematic uncertainties we find $\Omega_{\rm M}=0.352\pm0.017$ in flat-$\Lambda$CDM.  Supernova data alone now require acceleration ($q_0<0$ in $\Lambda$CDM) with over $5\sigma$ confidence.
We find $(\Omega_{\rm M},w)=(0.264^{+0.074}_{-0.096},-0.80^{+0.14}_{-0.16})$ in flat-$w$CDM.
For flat-$w_0w_a$CDM, we find $(\Omega_{\rm M},w_0,w_a)=(0.495^{+0.033}_{-0.043},-0.36^{+0.36}_{-0.30},-8.8^{+3.7}_{-4.5})$, consistent with a constant equation of state to within $\sim2\sigma$.
Including Planck CMB, SDSS BAO, and DES $3\times2$-point data gives $(\Omega_{\rm M},w)=(0.321\pm0.007,-0.941\pm0.026)$.
In all cases dark energy is consistent with a cosmological constant to within $\sim2\sigma$.  Systematic errors on cosmological parameters are subdominant compared to statistical errors; these results thus pave the way for future photometrically classified supernova analyses. \\ \\
{\em Keywords:} supernovae, cosmology, dark energy % Put the keywords here instead of in the keywords command that was causing spacing issues.
\end{abstract}
%\keywords{supernovae, cosmology, dark energy}


\section{Introduction} \label{sec:intro}
The standard cosmological model posits that the energy density of the Universe is dominated by dark components that have not been detected in terrestrial experiments and thus do not appear in the standard model of particle physics. Known as cold dark matter and dark energy, their study represents an opportunity to deepen our understanding of fundamental physics.

The Dark Energy Survey (DES) was conceived to characterize the properties of dark matter and dark energy with unprecedented precision and accuracy through four primary observational probes \citep{DES2005_design,bernstein12,DES2016_morethanDE,DESbook_2020}. One of these four probes is the Hubble diagram (redshift-distance relation) for Type Ia supernovae (SNe~Ia), which act as standardizable candles \citep{rust74,pskovskii77,phillips99} to constrain the history of the cosmic expansion rate. To implement this probe, the DES SN survey was designed to provide the largest, most homogeneous sample of high-redshift supernovae ever discovered. The two papers that first presented evidence for the accelerated expansion of the universe \citep{riess98,perlmutter99} used a total of 52 high-redshift supernovae with sparsely sampled light-curve measurements in one or two optical passbands. Building on two decades of subsequent improvements in SN surveys and analysis, we present here the cosmological constraints using the full 5-year DES SN dataset, consisting of well-sampled, precisely calibrated light curves for 1635 new high-redshift supernovae observed in four bands $g, r, i, z$.

For the last decade, SN~Ia cosmology constraints have largely come from combining data from many surveys.  The recent Pantheon+ analysis \citep{scolnic21_pantheonpsample, brout22_pantheon} combined three separate mid-$z$ samples ($0.1<z<1.0$), 11 different low-$z$ samples ($z<0.1$), and four separate high-$z$ samples ($z>1.0$), each with different photometric systems and selection functions \citep{Gilliland_1999,hicken09,Riess_2001,Riess_2004,Riess_2007,2011ApJ...737..102S,hicken12,Suzuki_2012,2013MNRAS.433.2240G,betoule14,krisciunas17,foley17,Riess_2018,2018PASP..130f4002S,DES_SMP, smithDAndrea20}. 
The DES sample, which rivals in number the entirety of Pantheon+, does not have the low-redshift ($z<0.1$) coverage to completely remove the need for external low-$z$ samples, but at higher redshift enables us to replace a heterogeneous mix of samples with a homogeneous sample of high quality, well-calibrated light curves. 

A key aim of the DES analysis was to minimize systematic (relative to statistical) errors to enable a robust analysis.  \citet{DES5yr_analysis} shows that our error budget is dominated by statistical uncertainty, in contrast to most SN cosmology analyses of the last decade, for which the systematic uncertainties equalled or exceeded the statistical uncertainties \citep{betoule14, scolnic18, DES-SN3YR}. We also highlight that the most critical sources of systematics are those related to the lack of a homogeneous and well calibrated low-$z$ sample.

As the DES sample enables a SN~Ia measurement of cosmological parameters that is largely independent of previous SN cosmology analyses, we have been careful to ``blind'' our analysis (see Sec.~\ref{sec:unblind}).  The analysis work described in \citet{DES5yr_analysis}, which stops just short of constraining cosmological parameters, was shared widely with the DES collaboration, evaluated, and approved before unblinding.  Unblinding standards included multiple validation checks with simulations and full accounting and explanation of the error budget. No elements of the analysis were changed after unblinding.  

\begin{figure}
    \includegraphics[width=\linewidth]{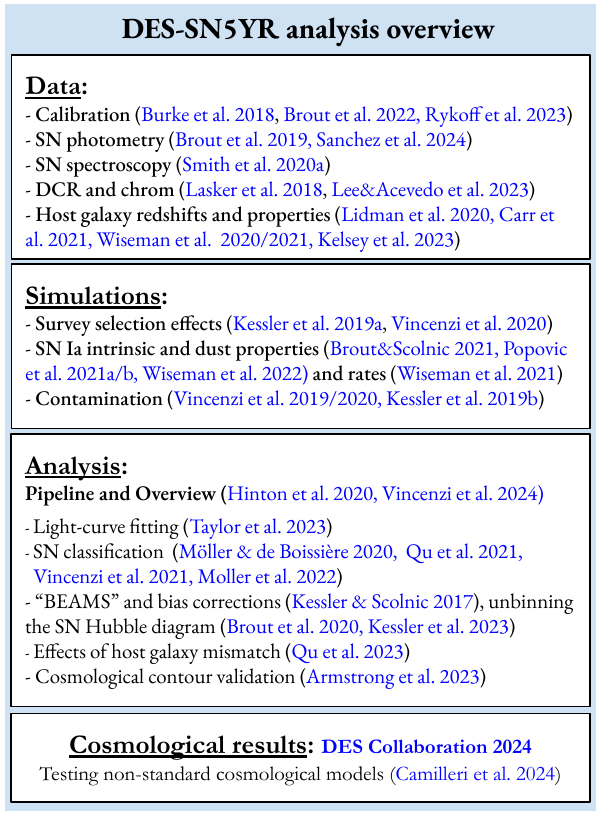}
    \caption{Overview of supporting papers for DES-SN5YR cosmological results.}
    \label{fig:overview}
\end{figure}

In this paper we review the analysis of the complete DES SN dataset (as detailed in many supporting papers; see Fig.~\ref{fig:overview}) and present the cosmological results.  An important advance on most previous analyses is that we use a photometrically classified rather than spectroscopically classified sample \citep{moller20,qu21}, and implement advanced techniques to classify SN Ia and incorporate classification probabilities in the cosmological parameter estimation \citep{kunz07,kunz12,hlozek12}.
While this advance increases the complexity of the analysis, in this work and previous papers \citep{vincenzi21, moller22} we show that the impact of non-SN~Ia contamination due to photometric misclassification is well below the statistical uncertainty on cosmological parameters, 
 and this constitutes one of the key results of our analysis.
 
Combining our DES data with a low-redshift sample (see Sec.~\ref{sec:data}), we fit the Hubble diagram to test the standard cosmological model as well as multiple common extensions including spatial curvature, non-vacuum dark energy, and dark energy with an evolving equation of state parameter. In \citet{camilleri24} we present fits to more exotic models. 

The structure of the paper is as follows.  We begin in Sec.~\ref{sec:data} by describing the dataset, its acquisition, reduction, calibration, and light-curve fitting.  
We summarize the models we test in Sec.~\ref{sec:models} before presenting the results in Sec.~\ref{sec:results}; our discussion and conclusions follow in Sec.~\ref{sec:discussion} and Sec.~\ref{sec:conclusions}. The details of our data release, which includes the code needed to reproduce our results, appear in \citet{sanchez24}. 

\begin{figure*}
    \centering
    \includegraphics[height=144mm]{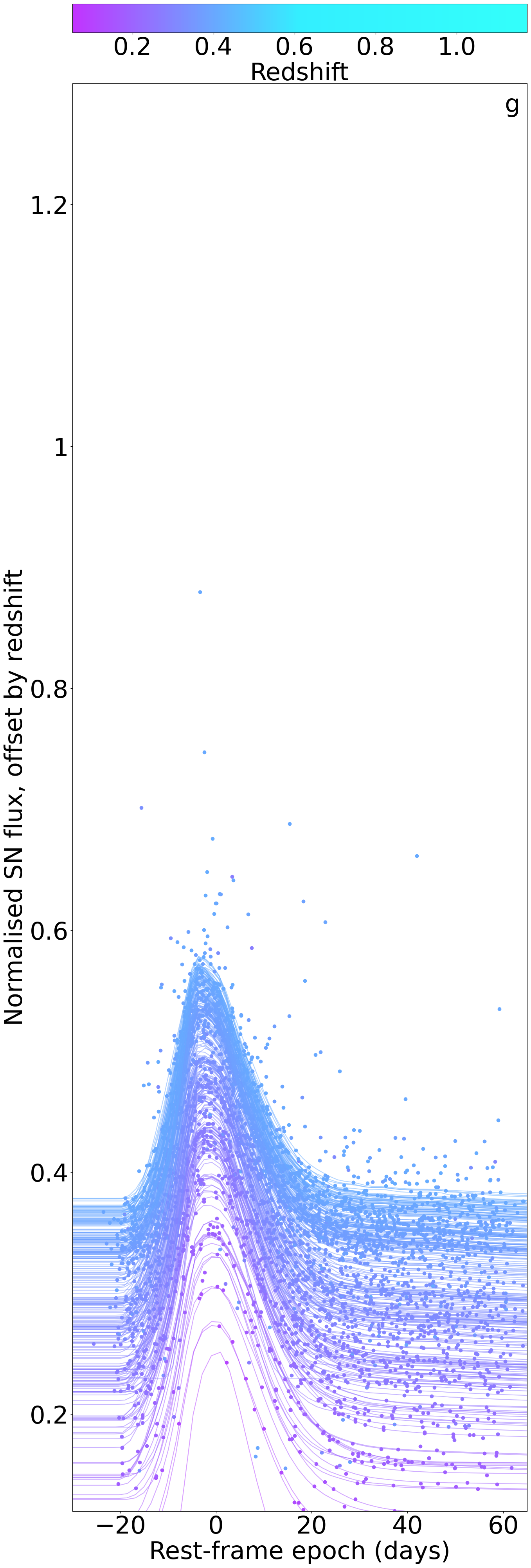}
    \includegraphics[height=144mm]{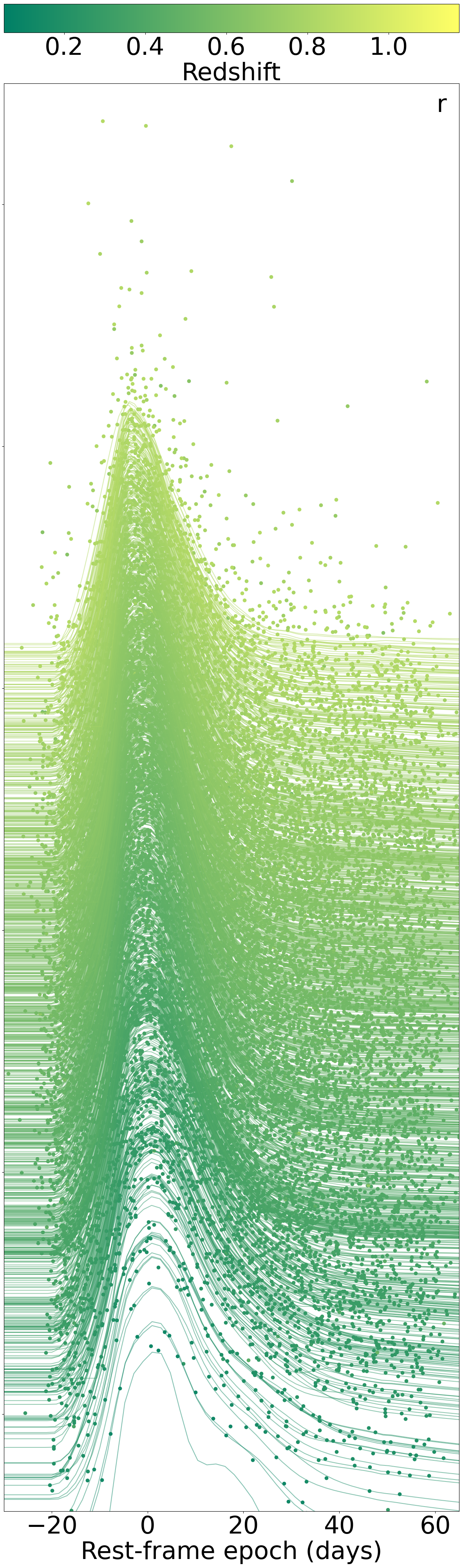}
    \includegraphics[height=144mm]{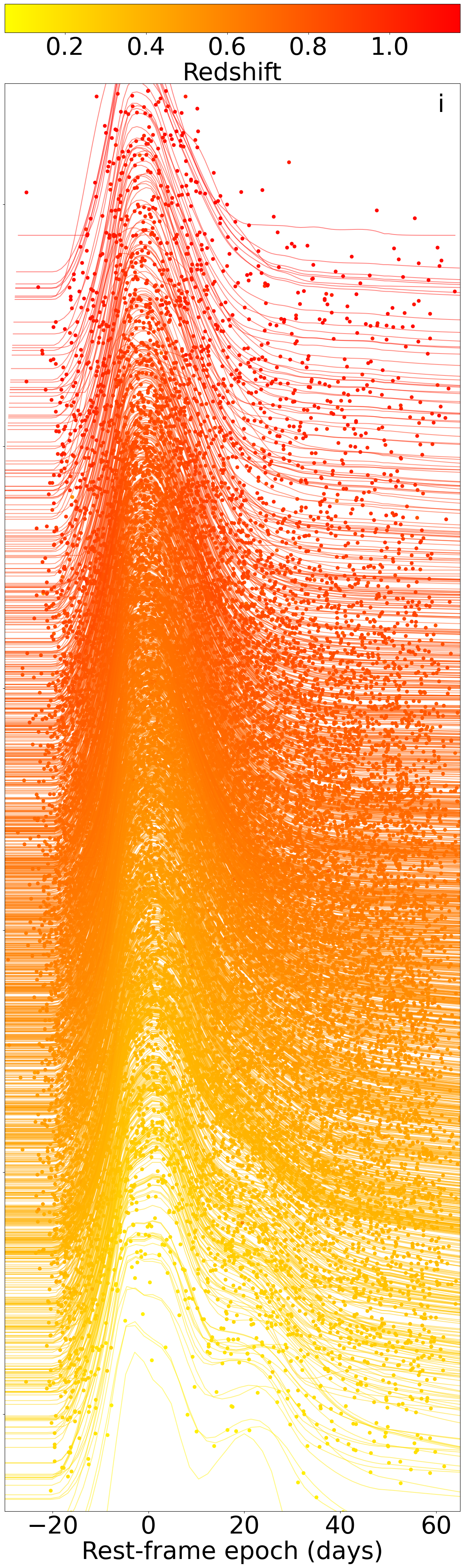}
    \includegraphics[height=144mm]{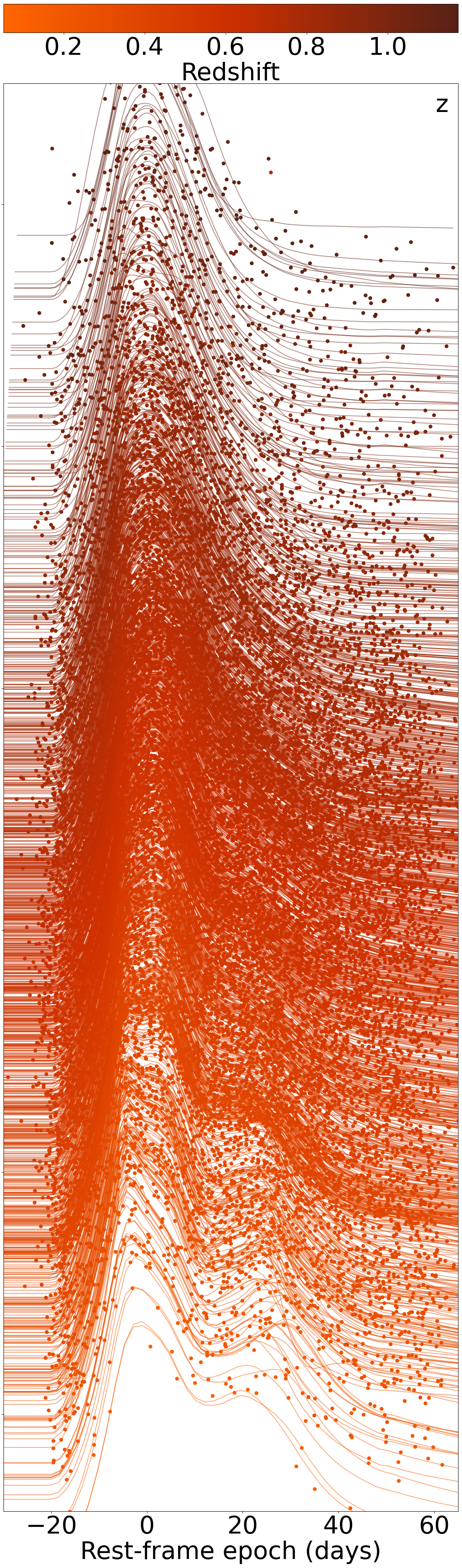}
    \caption{All DES light curves, showing observed magnitudes in $g$, $r$, $i$, and $z$ bands (left to right respectively) normalized by the maximum brightness of each light curve, and with the time-axis de-redshifted to the rest-frame.  Each light curve has been arbitrarily offset by their redshift, with higher-redshift objects higher on the plot (as labeled on vertical axis).  Lines show best-fit SALT3 light-curve fits. The $g$-band and $r$-band light curves are not used above $z\sim0.4$ and $z\sim0.85$ respectively because that corresponds to the redshifts at which the lower-wavelength limit of the SALT3 model ($3500\AA$ in the rest frame) passes out of their observed wavelength ranges.}
    \label{fig:lightcurves}
\end{figure*}

%%%%%%%%%%%%%%%%%%%%%%%%%%%%%%%%%%%%%%%%
\section{Data and Analysis}\label{sec:data}
\subsection{DES and Low-redshift SNe}
Our primary dataset is the full five years of DES SNe, which we combine with a historical set of nearby supernovae from CfA3 \citep{hicken09}, CfA4 \citep{hicken12}, CSP \citep[][DR3]{krisciunas17} and the Foundation SN sample \citep{foley17}. We refer to the combined DES plus historical dataset as \textbf{DES-SN5YR}. 

The DES supernova program was carried out over five seasons, August to February from 2013--2018, during which we observed ten $\sim 3\,{\rm deg}^2$ fields with approximately weekly cadence in four bands ($g, r, i, z$).  
Eight of the fields were observed to $5\sigma$ depth of ${\sim}23.5$~mag in all four bands (shallow fields) and two to a deeper limit of ${\sim}24.5$~mag (deep fields). 
See \cite{DECAM2015} for a summary of the Dark Energy Camera,
\citet{smithDAndrea20} for a summary of the supernova program, 
and \citet{diehl16,diehl18} for observational details.

The DES SNe were discovered via difference imaging \citep{kessler15} based on the method of \citet{alard98}. DES images are calibrated following the Forward Global Calibration Method \citep[FGCM;][]{burke18, sevilla-noarbe21,rykoff23}, and both DES and low-$z$ samples are recalibrated as part of the SuperCal-Fragilistic cross calibration effort described in \citet{brout22_fragilistic}. SN fluxes are determined using scene modeling photometry \citep{DES_SMP}; we include corrections from spectral energy distribution variations \citep{burke18,DES_chrom} and from differential chromatic refraction and wavelength-dependent seeing \citep*{DES5YR-DCR}. We estimate the overall accuracy of our calibrated photometry to be $\lesssim 5$~mmag. Host galaxies are assigned following the directional light radius (DLR) method \citep{sullivan06, gupta16, qu23_hostMismatch}, and host galaxy properties are determined as described by \citet{kelsey23} based on \citet{fioc99} using deep coadded images by \citet{DES_deepstacks}. Host galaxy spectroscopic redshifts are obtained primarily within the OzDES programme \citep{yuan15, childress17, lidman20}. The final data release of photometry of $\sim20,000$ candidates, redshifts of hosts, and host galaxy properties is presented in \citet{sanchez24}.

\begin{figure}
    \includegraphics[width=0.99\linewidth]{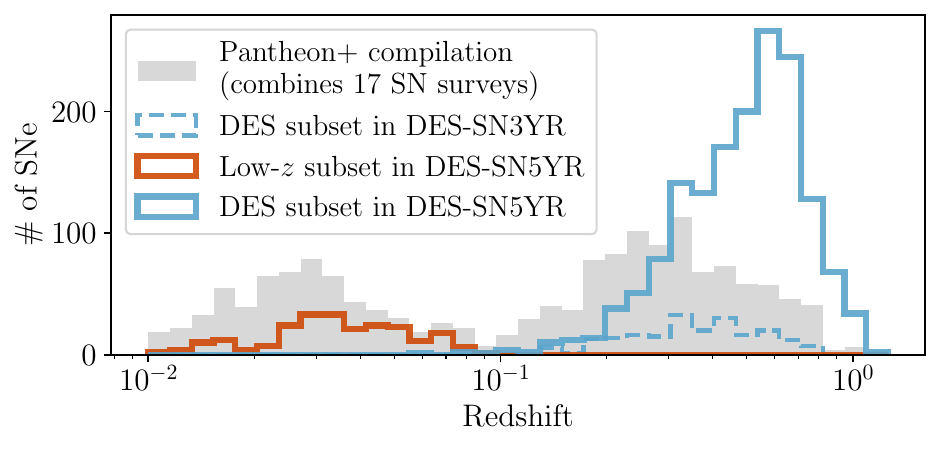}
    \caption{Histogram showing the redshift distribution of the DES-SN5YR sample, with new DES SNe in blue and our low-$z$ sample in red.  For comparison the distribution of redshifts in the existing Pantheon+ sample is shown in grey \citep{brout22_pantheon}, which also includes the DES SNe from the DES-SN3YR analysis (blue dashed line). The five-year DES sample contains $\sim4\times$ more supernovae above $z\sim0.4$ than the Pantheon+ compilation.}
    \label{fig:histogram}
\end{figure}

We apply strict quality cuts to this sample of candidates to select our final high-quality sample for the Hubble diagram. The same quality cuts were applied to both the low-$z$ sample and the DES supernovae. First, we require a spectroscopic redshift of the host galaxy, good light-curve coverage (at least two detections with SNR$>5$ in two different bands), and a well converged light-curve fit using the SALT3 model\footnote{The SALT3 model consists of a spectral flux density as a function of phase and wavelength for type Ia supernovae.  Its three components are: $M_0$ describing the mean SN light curve, $M_1$ describing the deviations from $M_0$ that are correlated with light-curve width, and $CL$ describing the color-dependence.  See Eq.~1 of \citep{taylor23}.} \citep[][]{kenworthy21,taylor23}; this reduces the DES sample size to 3621.  Additional requirements include light-curve parameters (stretch and colour) within normal range for SNe Ia, a well-constrained time of peak brightness (uncertainty less than 2 days), good SALT3 fit-probability, and valid distance-bias correction from our simulation \citep[see Table 4 of][for more detail]{DES5yr_analysis}.
Our final Hubble-diagram sample includes {\bf 1635} supernovae, of which 1499 have a machine-learning probability of being a Type Ia greater than 50\% (see Sec.~\ref{sec:analysis}). Note that we do not perform a cut on this machine-learning probability, rather we use it in the BEAMS formalism that produces our Hubble diagram and to weight the SN distance uncertainties in the fits to the final Hubble diagram \citep{kessler23_binning}.  The set of all DES light curves is visualised in Fig.~\ref{fig:lightcurves}.

Since we focus on minimizing potential systematic errors, we only use the best-calibrated, most homogeneous sample of low-$z$ SNe Ia.  To reduce the impact of peculiar velocity uncertainties we remove SNe with $z<0.025$.  We furthermore combine only a subset of the available low-redshift samples: CfA3\&4, CSP, and Foundation SNe, which are the four largest low-$z$ samples with the most well-understood photometric calibration. 
Our low-$z$ sample thus totals 194 SNe with $z<0.1$; this can be compared to Pantheon+, for which the low-$z$ sample was almost four times larger (741 SNe at $z<0.1$).  We have thus exchanged the statistical constraining power
of more low-$z$ SNe for better control of systematics. The redshift distribution of our sample compared to the compilation of historical samples in Pantheon+ is shown in Fig.~\ref{fig:histogram}. To conclude, the final DES-SN5YR sample includes 1635 DES SNe and 194 low-$z$ external SNe, for a total of \textbf{1829} SNe.

%%%%%%%%%%%%%%%%%%%%%%%%%%%%%%%%%%%%%%%%
\begin{figure*}
    \includegraphics[width=0.99\linewidth]{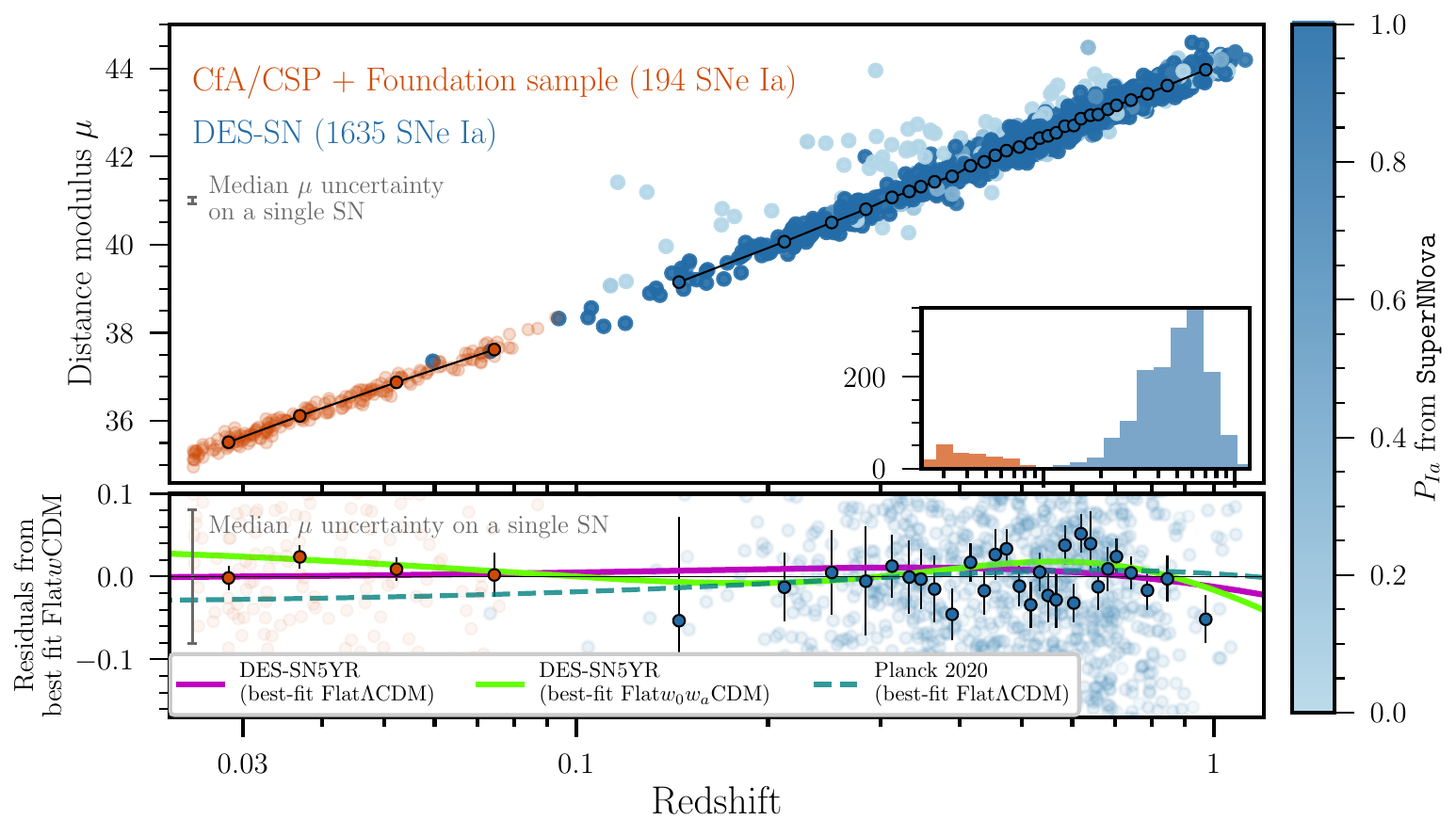}
    \caption{Hubble diagram of DES-SN5YR. We show both the single SN events and the redshift-binned SN distance moduli. Redshift bins are adjusted so that each bin has the same number of SNe ($\sim50$).  
    The \numdeshd\ new DES supernovae are in blue, and in the upper panel they are shaded by their probability of being a Type Ia; most outliers are likely contaminants (pale blue).  
    The inset shows the number of SNe as a function of redshift (same $z$-range as the main plot).  The lower panel shows the difference between the data and the best fit Flat-$w$CDM model from DES-SN5YR alone (third result in Table~\ref{tab:cosmo_results}), and overplots three other best fit cosmological models --- 
    Flat-$\Lambda$CDM model from DES-SN5YR alone (magenta line, first result in Table~\ref{tab:cosmo_results}), 
    Flat-$w_0w_a$CDM model from DES-SN5YR alone (green line, fourth result in Table~\ref{tab:cosmo_results}), 
    and Planck 2020 Flat-$\Lambda$CDM model without SN data (dashed line, $\om^{\rm Planck}=$\FlcdmPLANCKomegam). }
    \label{fig:hubblediagram}
\end{figure*}

\subsection{From light curves to Hubble diagram}\label{sec:analysis}
A critical step in the cosmology analysis is to convert each supernova's light curve (magnitude vs time in multiple bands; see examples in Fig.~\ref{fig:lightcurves}) to a single calibrated number representing its standardized magnitude and estimated distance modulus.  

To achieve this, we use the SALT3 light-curve fitting model as presented in \citet{kenworthy21,taylor23} and retrained in \citet{DES5yr_analysis} to determine the light-curve fit parameters,  amplitude of the SN flux ($x_0$), stretch ($x_1$), and color ($c$).
These fitted parameters are used to estimate the distance modulus, $\mu\equiv m-M$, using an adaptation of the Tripp equation \citep{tripp98} that includes a correction for observed correlations between SN~Ia luminosity and host properties, ${\gamma}G_{\rm host} =\pm\gamma/2$. Here $\gamma$ is the size of the step and $G_{\rm host}$ is the property of the host galaxy that is used to determine the step (i.e.\ mass or color); the sign is $+$ if $G_{\rm host}$ is above the step or $-$ if below.
This correction has historically been described as a ``mass step'' but we also consider the possibility that it is a ``color step'' \citep[see Sec.~2.2 of][]{DES5yr_analysis},  
\beq \mu_{{\rm obs},i} =m_{x,i} +\alpha x_{1,i} -\beta c_{i}+\gamma G_{{\rm host},i} -M - \mubias, \label{eq:tripp}\eeq
where $m_x=-2.5\log_{10}(x_0)$.\footnote{
Following \citet{marriner11}, we replace the traditional $m_B$ notation 
with $m_x$, because in the SALT2 and SALT3 models the amplitude term, $x_0$, is not related to any particular filter band.} %Traditionally, the $m_{\rm B}$ notation was used instead of the term $m_x$. However, in the SALT2 and SALT3 models the light-curve amplitude is parameterized by the amplitude term $x_0=10^{-m_B^{\prime}/2.5}$ plus an offset that makes $m_B^{\prime}$ close to the magnitude in the B-band. This updated formalism was introduced by \citet{marriner11}.} 
The constants $\alpha$, $\beta$, and $\gamma$ are global parameters determined from the likelihood analysis of all the SNe on the Hubble diagram, while the terms subscripted by $i$ refer to parameters of individual SNe. We find $\alpha=0.161\pm0.001$, $\beta=3.12\pm0.03$, and $\gamma=0.038\pm0.007$.  
We marginalize over the absolute magnitude $M$ (see Sec.~\ref{sec:models}). The final term in Eq.~\ref{eq:tripp} accounts for selection effects, Malmquist bias, and light curve fitting bias.

The nuisance parameters and $\mubias$ term in Eq.~\ref{eq:tripp} are determined using the ``BEAMS with Bias Corrections'' (BBC) framework \citep{Kessler_2017}. 
In particular, bias corrections $\mubias$ are estimated from a large simulation of our sample. The simulation models the 
rest-frame SN~Ia spectral energy distribution (SED) at all phases, 
SN correlations with host-galaxy properties,
SED reddening through an expanding universe, 
broadband $griz$ fluxes,
and instrumental noise \citep[see Fig.~1 in][]{kessler19_DESbiascor}.  
Using Eq.~\ref{eq:tripp} there remains intrinsic scatter of $\sim 0.1$ mag in Hubble residuals. 
Following the numerous recent studies on  understanding and modelling SN~Ia dust extinction and progenitors \citep{wiseman2021,wiseman22,2022arXiv221114291D, DES:2022hav, DES:2022tjd, 2023MNRAS.518.1985M}, we model this residual scatter using the dust-based model from \citet{brout21_BS20} [BS21]; \citet{popovic21_dust2dust}.
In contrast to previously used models in K13, the BS21 model accurately models the Hubble residual bias and scatter as a function of the fitted SALT2 color (see Fig.~5 in \citet{DES5yr_analysis}, and Fig.~6 in \citet{brout21_BS20}). Due to uncertainties in the fitted dust parameters \citep{popovic21_dust2dust}, this intrinsic scatter model remains the largest source of systematic uncertainty from the simulation.

As we do not spectroscopically classify the SNe and thus expect contamination from core-collapse (CC) supernovae, we perform machine learning light-curve classification on the sample following \citet{vincenzi21, moller22}. We implement two advanced machine learning classifiers, SuperNNova \citep{moller20} and SCONE \citep{qu21} and use state-of-the-art simulations to model contamination \citep[estimated to be $\sim6.5$\%, see Table~10 and Sec.~7.1.5 of][]{DES5yr_analysis}. Classifiers are trained using core-collapse and peculiar SN~Ia simulations based on \citet{Vincenzi_2020} and using state-of-the-art SED templates by \citet{Vincenzi_2019, 2019PASP..131i4501K}. These DES simulations are the first to robustly reproduce the contamination observed in the 
Hubble residuals \citep[][Table 10]{Vincenzi_2020, DES5yr_analysis}.

For each SN, the trained classifiers assign a probability of being a Type Ia, and these probabilities are included within the BEAMS framework to marginalize over core-collapse contamination and produce the final Hubble Diagram \citep{kunz12, hlozek12}. The final DES-SN5YR Hubble diagram is shown in Fig.~\ref{fig:hubblediagram} and includes 1829 SNe. 

As discussed in \citet{kessler23_binning, DES5yr_analysis}, the probability that each supernova is a Type Ia ($P_{\rm Ia}$) is incorporated in the BBC fit and used to calculate a BEAMS probability, $\PBIa$ \citep[see Eq.~9 in ][]{kessler23_binning}. BEAMS probabilities are used to inflate distance uncertainties of likely contaminants by a factor $\propto 1/\sqrt{\PBIa}$ \citep[see Eq.~10 in][]{DES5yr_analysis}. 
Therefore, the \textbf{released Hubble diagram data includes distance bias corrections and inflated distance uncertainties} (see App.~\ref{sec:release}), enabling users to fit the Hubble diagram without applying additional corrections.
With this BEAMS uncertainty weight, we find 75 SNe with distance modulus uncertainties $\sigma_{\mu,i,\mathrm{final}}>1$~mag and 1331 SNe with $\sigma_{\mu,i,\mathrm{final}}<0.2$~mag.\footnote{Applying a binary classification-based cut (SN~Ia or not) is not optimal, as it assumes the classification is perfect. However, we test the binary-cut-based approach by using only the 1499 SNe classified with $P_{\rm Ia}>0.5$ and assuming they are a pure SN Ia sample. We show that the measured shift in $w$ is 
small compared to the statistical uncertainties \citep[Table~11 of][]{DES5yr_analysis}.}

\citet{DES5yr_analysis} stops short of performing cosmological constraints but provides the corrected distance moduli $\mu$ along with their uncertainties $\sigma_{\mu}$, redshifts for each SN, and a statistical+systematic covariance matrix $C$, which we describe further in Sec.~\ref{sec:models}.

\citet{armstrong23} presents validation of the cosmological contours produced by our pipeline. Validation that our analysis pipeline is insensitive to the cosmological model assumed in our bias correction simulation appears in \citet{camilleri24}. 

%%%%%%%%%%%%%%%%%%%%%%%%%%%%%%%%%%%%%%%%
\begin{table*}[]
    \centering
    \begin{tabular}{r|l|l}
    \hline\hline
        {\bf Cosmological Model} & Friedmann Equation: $\mathbf{E(z)=H(z)/H_0=}$ & {\bf Fit Parameters $\Theta$} \\ \hline\vspace{2mm}
        Flat-$\Lambda$CDM & $\left[\om (1+z)^3 + (1-\om)\right]^{1/2}$ & $\om$ \\
        $\Lambda$CDM & $\left[\om (1+z)^3 + \oll + (1-\om-\oll)(1+z)^{2}\right]^{1/2}$ & $\om, \oll$ \\
        Flat-$w$CDM & $\left[\om (1+z)^3 + (1-\om)(1+z)^{3(1+w)}\right]^{1/2}$ & $\om, w$ \\
        %$w$CDM & $\om (1+z)^3 + \oll (1+z)^{3(1+w)}+ (1-\om-\oll)(1+z)^{2}$ & $\om, \oll, w$ \\
        Flat-$w_0$$w_a$CDM & $\left[\om (1+z)^3 + (1-\om) (1+z)^{3(1+w_0+w_a)}e^{-3w_a z/(1+z)}\right]^{1/2}$ & $\om, w_0, w_a$ \\
        \hline
    \end{tabular}
    \caption{Variations on the standard cosmological model that are tested in this paper, their Friedmann Equations, and the free parameters in the fit.  }
    \label{tab:models}
\end{table*}

\vspace{5mm}
\subsection{Unblinding criteria}\label{sec:unblind}
Throughout our analysis, cosmological parameters estimated from \textit{real data} were blinded. We validate our entire pipeline on detailed catalogue-level simulations and examine the cosmological parameters estimated from \textit{simulations} to test that the input cosmology is recovered. In addition to the many tests described in \citet{DES5yr_analysis}, the final unblinding criteria that our data passed were:
\begin{itemize}
\item {\bf Accuracy of simulations:} Reduced $\chi^2$ between the distribution of data and simulations across a variety of observables (redshift, SALT3 parameters and goodness of the fit, maximum signal-to-noise ratio at peak, host stellar mass) is required to be between 0.7 and 3.0 \citep[see][Fig.~3-4]{DES5yr_analysis}.
\item {\bf Pipeline validation using DES simulations:} Demonstrate that our pipeline recovers the input cosmology. We produce 25 data-size simulated samples (statistically independent) assuming a Flat-$\Lambda$CDM universe with best-fit Planck value of $\om$ and analyze them the same way as real data. We fit each Hubble diagram assuming a Flat-$w$CDM model with a Planck prior and find a mean bias of $w-w_{\rm true} \simeq 0.001 \pm 0.020$, where $w$ is the mean value of the marginalized posterior of the dark energy equation of state parameter over the 25 samples, and $w_{\rm true}=-1$ is the model value of that parameter input to the simulation.
\item {\bf Validation of contours:} ensuring that our uncertainty limits accurately represent the likelihood of the models \citep{armstrong23}.
\item {\bf Independence of reference cosmology:} ensuring that our results are sufficiently independent of cosmological assumptions that enter our bias correction simulations \citep{camilleri24}.
\end{itemize}

\subsection{Combining SN with other cosmological probes}
\label{sec:other probes}
We combine the DES-SN5YR cosmological constraints with measurements from other complementary cosmological probes. In particular, we use:
\begin{itemize}
    \item{Cosmic Microwave Background (CMB) measurements of the temperature and polarisation power spectra (TTTEEE) presented by the \citet{planck18_VI}. We use the Python implementation of Planck’s 2015 \texttt{Plik\_lite} \citep{prince19}.}
    \item{Weak lensing and galaxy clustering measurements from the DES3$\times$2pt year-3 magnitude-limited (MagLim) lens sample; %(referred to as DES Y3 3$\times$2pt); 
    $3\times2$-point refers to the simultaneous fit of three 2-point correlation functions, namely galaxy-galaxy, galaxy-lensing, and lensing-lensing correlations \citep{DES3x2_2022,DES3x2extensions_2023}.}
    \item{Baryon acoustic oscillation (BAO) measurements as presented in the extended Baryon Oscillation Spectroscopic Survey paper \citep[eBOSS;][]{dawson16,alam21}, which adds the BAO results from SDSS-IV \citep{blanton17} to earlier SDSS BAO data. Specifically, we use ``BAO'' to refer to the BAO-only measurements from the Main Galaxy Sample \citep{ross15}, BOSS \citep[SDSS-III][]{alam17}, eBOSS LRG \citep{bautista21}, eBOSS ELG \citep{demattia21}, eBOSS QSO \citep{hou21}, and eBOSS Lya \citep{bourboux20}.}
\end{itemize}

When combining these data we run simultaneous MCMC fits of the relevant data vectors.  We present three combinations: the simplest CMB-dependent combination CMB+SN, a CMB-independent combination BAO+3$\times$2pt+SN, and a combination of them all.

\section{Models and theory}\label{sec:models}
We present cosmological results for the standard cosmological model -- flat space with cold dark matter and a cosmological constant (Flat-$\Lambda$CDM) -- and some basic extensions, such as relaxing the assumption of spatial flatness ($\Lambda$CDM), allowing for constant equation of state parameter ($w$) of dark energy (Flat-$w$CDM), and including a linear parameterisation for time-varying dark energy (Flat-$w_0w_a$CDM) in which the equation of state parameter is given by $w=w_0+w_a(1-a)$ \citep{chevallier01,linder03}.  

To calculate the theoretical distance as a function of redshift we begin with the comoving distance,
\beq R_0\chi(\bar{z}) = \frac{c}{H_0}\int_0^{\zr} \frac{dz}{E(z)},  \label{eq:R0X} \eeq
where $\zr$ is the redshift due to the expansion of the Universe, $E(z)\equiv H(z)/H_0$ is the normalized redshift-dependent expansion rate and is given for each cosmological model by the expression in  Table~\ref{tab:models}, $R_0=c/(H_0\sqrt{|\ok|})$ is the scale factor with dimensions of distance (where subscript $0$ indicates its value at the present day), 
and $\ok\equiv1-\om-\oll$ is the curvature term.
The dimensionless scale factor ($a\equiv R/R_0$) at the time of emission for an object with cosmological redshift $\bar{z}$ is $a=1/(1+\bar{z})$.
The luminosity distance is given by, \beq \dl(z_{\rm obs},\bar{z}) = (1+z_{\rm obs})R_0 S_k(\chi(\bar{z})), \eeq
where $z_{\rm obs}$ is the observed redshift, and the curvature is captured by $S_k(\chi)=\sin\chi$, $\chi$, and $\sinh\chi$ for closed ($\ok<0$), flat ($\ok=0$), and open ($\ok>0$)  universes respectively.\footnote{When $\ok=0$ the term $R_0S_k(\chi)$ becomes $R_0\chi$ and can be calculated directly from Eq.~\ref{eq:R0X}, bypassing the infinite $R_0$.}   

To compare data (Eq.~\ref{eq:tripp}) to theory we calculate the theoretical distance modulus, which is dependent on the set of cosmological parameters we are interested in ($\Theta$, given in the right column of Table~\ref{tab:models}),
\beq \mu(z,\Theta) = 5 \log_{10}(\dl (z,\Theta)/1~{\rm Mpc}) + 25.\label{eq:mu} \eeq

We compute the difference between data and theory for every $i$th supernova, $\Delta \mu_i = \mu_{{\rm obs},i} - \mu(z_i,\Theta)$, and find the minimum of
\beq \chi^2 = %\sum_{ij}^{N_{\rm SNe}} 
\Delta \mu_i \mathcal{C}_{ij}^{-1} \Delta \mu_j^T \label{eq:chi2}~,\eeq 
where $\mathcal{C}^{-1}$ is the inverse covariance matrix (including both statistical and systematic errors) of the $\Delta \mu$ vector \citep[see Sec.~3.6 of][]{DES5yr_analysis}. 

The uncertainty covariance matrix includes a diagonal statistical term (discussed Sec.~\ref{sec:analysis}) and a systematic term.
The systematic covariance matrix is built following the approach in \cite{conley11} and accounts for systematics such as calibration, intrinsic scatter, and redshift corrections \citep[see Table 6 of][]{DES5yr_analysis}.  Each element of the covariance matrix expresses the covariance between two of the SNe in the sample.  The covariance matrix has dimensions of the number of supernovae $N_{\rm SNe}\times N_{\rm SNe}$ and we follow the formalism introduced by \citet{brout20_binning} and \citet{kessler23_binning}.

Finally, the absolute magnitude of SNe~Ia ($M$) and the $H_0$ parameter (which appears in the luminosity distance) are completely degenerate and therefore they are combined in the single parameter $\mathcal{M}=M+5\log_{10}(c/H_0)$.
All of our cosmology results are marginalized over this term. 
Therefore, the value of $H_0$ has no impact on the fitting of our cosmological results, and we do not constrain $H_0$.  While $\mathcal{M}$ has no impact on cosmology fitting, a precise value is needed to simulate bias corrections. The $\mathcal{M}$ uncertainty is below 0.01, resulting in a negligible impact on bias corrections \citep{brout22_pantheon,camilleri24}.

%%%%%%%%%%%%%%%%%%%%%%%%%%%%%%%%%%%%%%%%
\section{Results}\label{sec:results}

With the new DES high-redshift supernova sample we can put strong constraints on cosmological models.  Of particular interest is whether dark energy is consistent with a cosmological constant or whether its density and/or equation of state parameter varies over the wide redshift range of our sample. 
The results of our cosmological fits are outlined in this section and summarized in Table~\ref{tab:cosmo_results}, and their implications are explored in Sec.~\ref{sec:discussion}.

We estimate cosmological constraints using Markov Chain Monte Carlo (MCMC) methods as implemented in the CosmoSIS framework \citep{COSMOSIS}, 
the samplers \texttt{emcee} for best fits \citep{emcee}, and PolyChord for tension metrics \citep{polychord},\footnote{For each \texttt{emcee} fit we use a number of walkers that is at least twice the number of parameters and ensure the number of samples in the chain is greater than 50 times the autocorrelation function, $\tau$ ($N_{\rm samples}/\tau>50$). For each PolyChord fit, we use a minimum of 60 live points, 30 repeats, and an evidence tolerance requirement of 0.1 (except for $\Lambda$CDM with all datasets combined, for which we accepted a slightly weaker tolerance because convergence was too slow). 
When combining with other datasets we run simultanous MCMC chains including all relevant data vectors. Flat priors that  encapsulate at least the 99.7\% confidence region were chosen in each case, and we summarise those priors in Appendix~\ref{sec:priors}.} except for fits that include \BAOpt, which are calculated using PolyChord for both best fit and tensions.\footnote{The main advantage of \texttt{emcee} is it gives slightly more accurate best fit $\chi^2$ than PolyChord.  However, we decided the tiny improvement in accuracy was not worth the environmental impact \citep{stevens20} of the extra compute time (which was substantial for the many-dataset fits).}
For all fits we present the median of the marginalized posterior 
and cumulative 68.27\% confidence intervals. The chains and code (with the flexibility to test other statistical choices) are publicly available (see Appendix~\ref{sec:release}). 
Figs.~\ref{fig:FlatLCDM}, \ref{fig:LCDM}, \ref{fig:FlatwCDM} and \ref{fig:Flatw0waCDM} all present the joint probability contours for 68.3\% and 95.5\%.

\subsection{Constraints on Cosmological Parameters}\label{sec:constraints}
\begin{figure}
    \centering
    \includegraphics[width=0.90\linewidth]{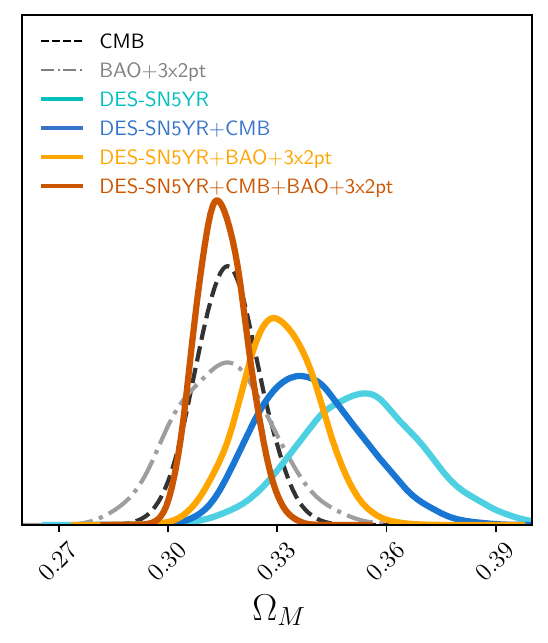}
    \caption{Constraints on matter density in the Flat-$\Lambda$CDM model from DES-SN5YR only (cyan), DES-SN5YR combined with CMB constraints from \citet{planck18_VI} (blue), and DES-SN5YR combined with BAO+3$\times$2pt (orange), and all probes combined (DES-SN5YR+BAO+3$\times$2pt and CMB constraints, red). CMB constraints only and BAO+3$\times$2pt constraints alone are also shown for comparison (dashed and dotted-dashed respectively).}
    \label{fig:FlatLCDM}
\end{figure}

\begin{figure}
    \centering
    \includegraphics[width=0.49\textwidth]{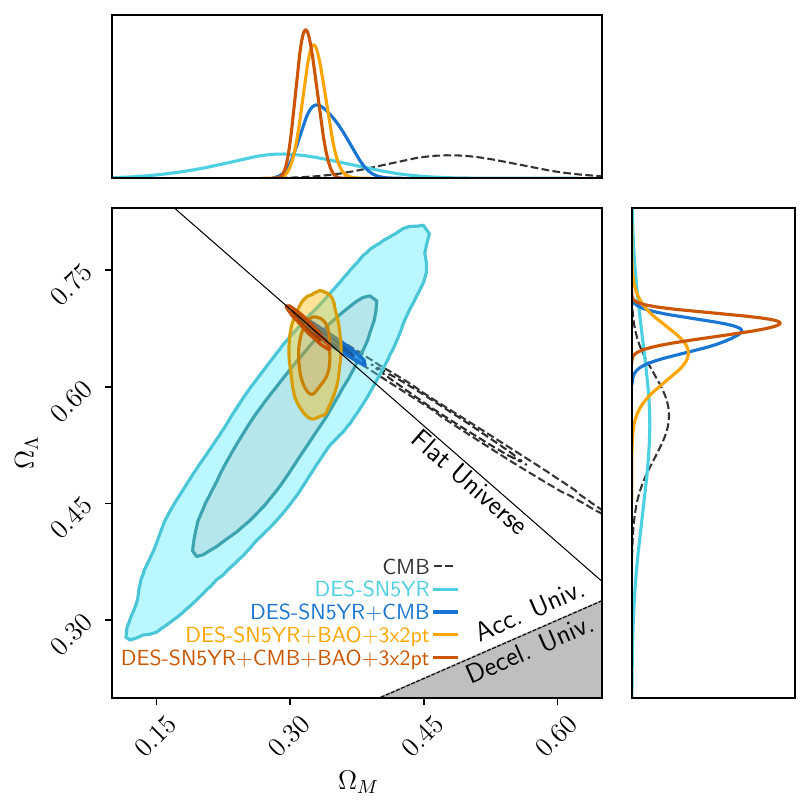}
    \caption{Constraints for $\Lambda$CDM model (non-zero curvature allowed) from the DES-SN5YR dataset only (cyan), from DES-SN5YR combined with BAO+3$\times$2pt (orange), from DES-SN5YR combined with CMB measurements (blue), and from all these combined (red). For comparison, we also present cosmological constraints from \citet{planck18_VI} only (black dashed). 
    }
    \label{fig:LCDM}
\end{figure}

\vspace{5mm}
\subsubsection{Flat-$\Lambda$CDM}\vspace{-3mm}
For the simplest parameterization, Flat-$\Lambda$CDM, $\om$ is the only free parameter.  We show the probability density function (PDF) of this constraint for DES-SN5YR %compared to \citet{planck18_VI} 
in Fig.~\ref{fig:FlatLCDM}; we measure a value of $\om=$\mflcdmSN.  We also show the probability distribution of  the \citet{planck18_VI} measurement of $\om^{\rm Planck}=$\FlcdmPLANCKomegam.  These are approximately\footnote{The distribution of points around the Hubble diagram is not perfectly Gaussian, as it is skewed due to lensing magnification and non-SN-Ia contamination.  This means the $\sigma$ values (especially at high-$\sigma$) are only approximate.} $2\sigma$ apart, but not in significant tension as discussed in Sec~\ref{sec:tensions}. 

Combining DES-SN5YR with Planck CMB gives $\om=$\mflcdmSNP, while combining with BAO+3$\times$2pt gives $\om=$\mflcdmSNBpt. Combining all three gives $\om=$\mflcdmSNPBpt.  Interestingly, the combination of all data sets (red in Fig.~\ref{fig:FlatLCDM}) gives a lower $\om$ than any of the other combinations.  
The reason can be seen in Fig.~\ref{fig:LCDM}, 
where all constraints cross the Flat Universe line to the upper left of any individual best fit. 
 % Difference is 2.09 sigma assuming Gaussian, which comes from (0.344-0.3166)/np.sqrt(0.0084**2+0.010**2)

\begin{figure}
    \centering
    \includegraphics[width=0.48\textwidth]{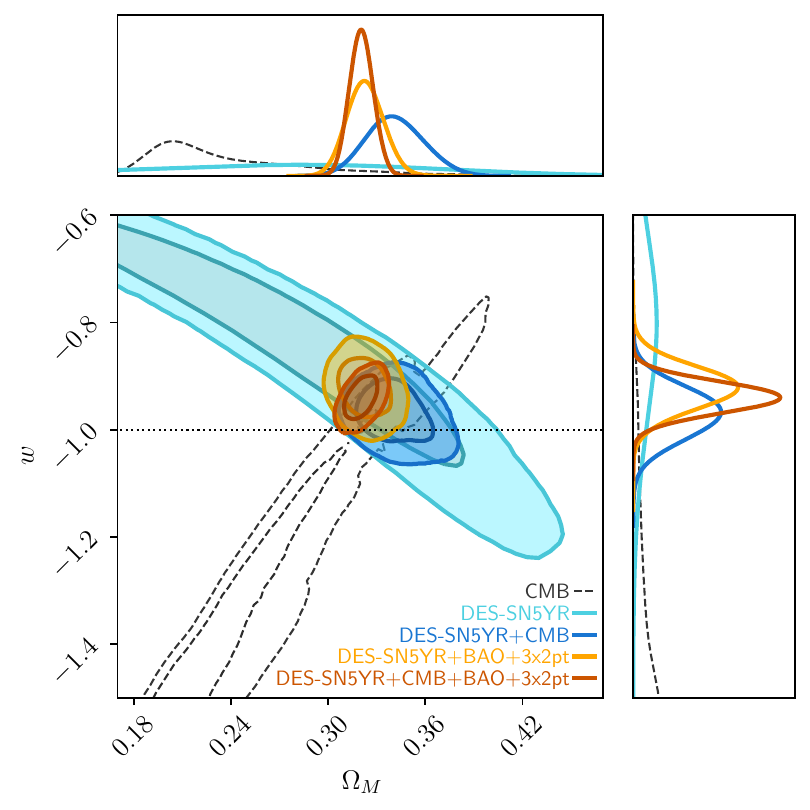}
    \caption{Same as Fig.~\ref{fig:LCDM} but for the Flat $w$CDM model. The horizontal dotted line marks the equation of state values for a cosmological constant, i.e. $w=-1$.} 
    \label{fig:FlatwCDM}
\end{figure}

\begin{figure*}
    \centering
    \includegraphics[width=0.9\textwidth]{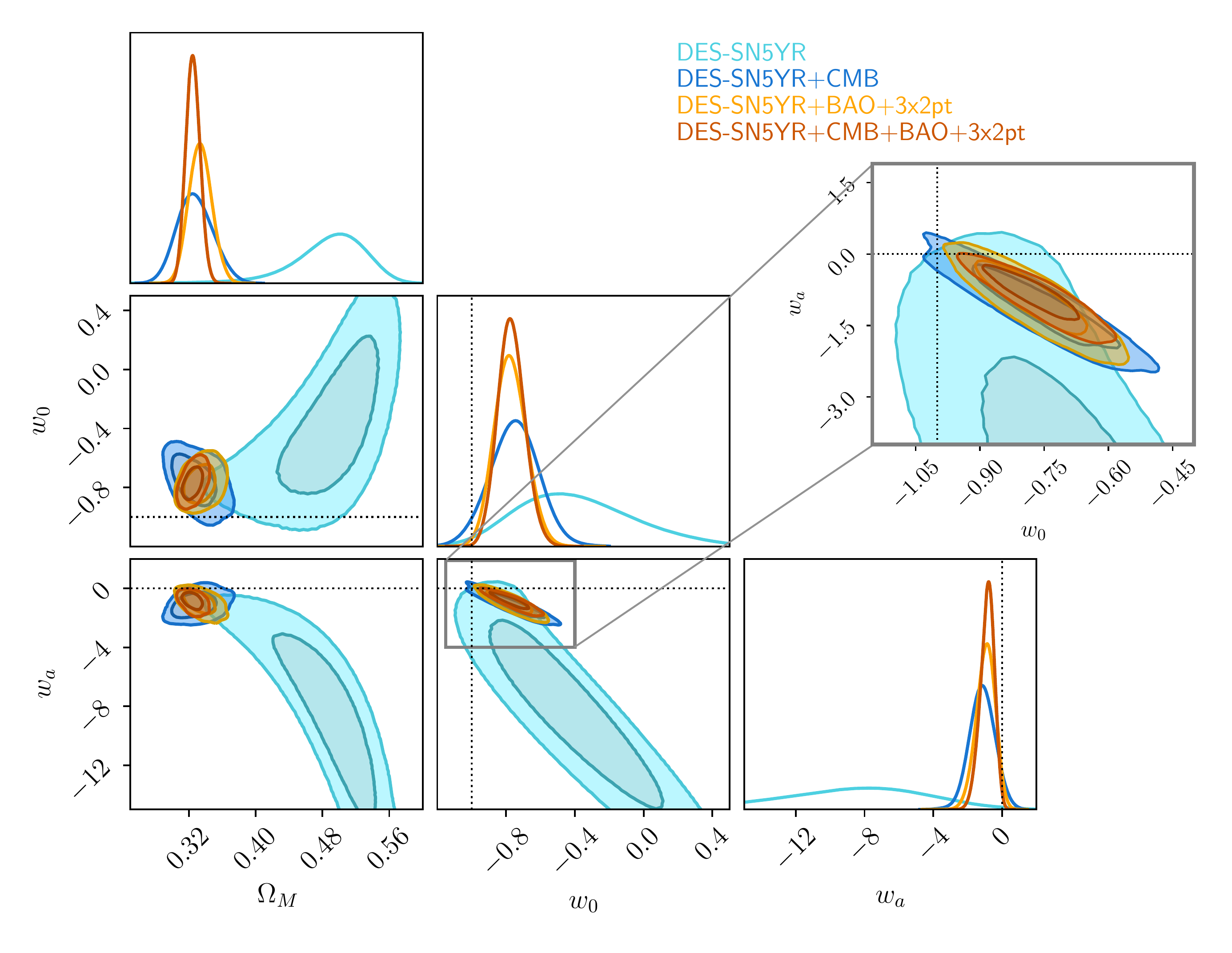}
    \caption{Same as Fig.~\ref{fig:LCDM} but for the Flat-$w_0 w_a$CDM model. The dashed crosshairs mark the equation of state values for a cosmological constant, i.e. $(w_0,w_a)=(-1,0)$. The residuals between the DES-SN5YR best fit Flat-$w_0w_a$CDM w.r.t. the Flat-$w$CDM model are presented in Fig.~\ref{fig:hubblediagram}.}
    \label{fig:Flatw0waCDM}
\end{figure*}

\subsubsection{$\Lambda$CDM}\label{sec:lcdm}
Fitting DES-SN5YR to the $\Lambda$CDM model,
we find  $(\om,\oll)$=(\mlcdmSN, \llcdmSN), consistent with a flat universe ($\ok$=\klcdmSN); see Fig.~\ref{fig:LCDM}. 
Combining DES-SN5YR with \BAOpt\ is also consistent with a flat Universe, with uncertainties on $\ok$ reduced to $\sim\pm0.034$, while the combination with Planck gives $\ok=$\klcdmSNP. The combination of all three gives $\ok=$\klcdmSNPBpt.

\subsubsection{Flat-$w$CDM}\label{sec:flatwdcm}
Fitting DES-SN5YR to the Flat-$w$CDM model,
we measure $(\om,w)=($\mfwcdmSN,\wfwcdmSN$)$; see Fig.~\ref{fig:FlatwCDM}.  This is consistent with a cosmological constant (within $2\sigma$), although our data favors a $w$-value that is slightly larger than $-1$.

The $w-\om$ contours from SN alone are highly non-Gaussian with a curved `banana'-shaped degeneracy.  
The best fit value for $w$ or $\om$ is thus an insufficient summary of the SN information, as a small shift along the degeneracy direction can result in large shifts in the best-fit values.  
To address this issue, in \citet{camilleri24} we introduce a new parameter, $\mate(z) \equiv -\ddot{a}/(aH_0^2)\equiv q(H/H_0)^2$. This combination of the deceleration parameter $q$ and the Friedmann equation $H/H_0$ follows the curve of the degeneracy in the $w-\om$ plane. Therefore, measuring $\mate(z)$ summarizes the supernova information in a single, almost degeneracy-free value.\footnote{Similar to the $S_8$ parameter used in lensing studies to approximate $\sigma_8$-$\om$ constraints.}   One has to choose the redshift at which one quotes $\mate(z)$, to best match the angle of the degeneracy for the redshift range of the sample. We find $\mate(z=0.2)=-0.340\pm0.032$ using DES-SN5YR only \citep[see][]{camilleri24}. 
This $Q_H$ value can be used to roughly approximate the DES-SN5YR results and characterize the constraining power without the need for a full fit to the Hubble diagram.

The degeneracy in the $w-\om$ plane is broken by combining SNe with external probes. 
Combining with Planck, we measure $(\om,w)=($\mfwcdmSNP,\wfwcdmSNP$)$, again within 2$\sigma$ of a cosmological constant.  Planck alone provides only a loose constraint on the equation of state parameter of dark energy, $w^{\rm Planck}=-1.51^{+0.27}_{-0.18}$; combining with DES-SN5YR reduces the uncertainty significantly due to the different degeneracy direction, demonstrating the combined constraining power of these two complementary probes.

Combining DES-SN5YR with \BAOpt\ we find $w=$\wfwcdmSNBpt, slightly over $2\sigma$ from the cosmological constant. 
This data combination demonstrates that these late-universe probes alone provide constraints that are consistent with -- and of comparable constraining power to -- the combination of SN and CMB data. 
The full combination of all data sets gives $w=$\wfwcdmSNPBpt.

\subsubsection{Flat-$w_0w_a$CDM}
Fitting DES-SN5YR alone to the Flat-$w_0w_a$CDM model
%The best-fit Flat-$w_0w_a$CDM model from DES-SN5YR alone 
gives an equation of state that is slightly over $2\sigma$ from a cosmological constant, marginally preferring a time-varying dark energy $(\om,w_0,w_a)=$(\mfwaSN, \wfwaSN, \wafwaSN$)$; see Fig.~\ref{fig:Flatw0waCDM}.  %(For results using a pivot redshift ($w_p$) see \citet{camilleri24}.) 

Combining DES-SN5YR and the CMB, we find $(\om,w_0,w_a)=$(\mfwaSNP,\wfwaSNP,\wafwaSNP), which again deviates slightly from the cosmological constant. The same trend is seen when combining with \BAOpt\ and with all data combined.  
The negative $w_a$ means that the dark energy equation of state parameter is {\em increasing} with time 
(sometimes referred to as a ``thawing'' model).
%and the dark energy density is decreasing with time.

\begin{table*}
    \centering
    \caption{Results for four different cosmological models, sorted into sections for different combinations of observational constraints. These are the medians of the marginalized posterior with 68.27\% integrated uncertainties (`cumulative' option in ChainConsumer). For each fit we also show the $\chi^2$ per degree of freedom as a measure of the goodness of fit.}
    \label{tab:cosmo_results}
    \begin{tabular}{lccccc}
        \hline
        & $\om$ & $\ok$ & $w_0$ & $w_a$ & $\chi^2/{\rm dof}$ \\ 
        
        \hline
        \multicolumn{6}{l}{\textbf{DES-SN5YR (no external priors)}}  \\
        \hline
		Flat-$\Lambda$CDM & \mflcdmSN  &  - & - & - & 1649/1734=0.951 \\ 
		$\Lambda$CDM & \mlcdmSN & \klcdmSN & - & - & 1648/1733=0.951 \\ 
		Flat-$w$CDM & \mfwcdmSN & - & \wfwcdmSN & - & 1648/1733=0.951 \\ 
    	Flat-$w_0w_a$CDM & \mfwaSN & - & \wfwaSN & \wafwaSN & 1641/1732=0.948 \\ 
        \hline
        \multicolumn{6}{l}{\textbf{DES-SN5YR + Planck 2020}}   \\
        \hline
        {Flat-$\Lambda$CDM} &  \mflcdmSNP &  - & - & - & 2237/2349=0.952 \\ 
		$\Lambda$CDM &  \mlcdmSNP &  \klcdmSNP & - & - & 2231/2348=0.950 \\ 
  		Flat-$w$CDM &  \mfwcdmSNP & - &  \wfwcdmSNP & - & 2234/2348=0.951 \\ 
  		Flat-$w_0w_a$CDM &  \mfwaSNP & - &  \wfwaSNP &  \wafwaSNP & 2231/2347=0.951 \\ 
		\hline
        \multicolumn{6}{l}{\textbf{DES-SN5YR + SDSS BAO and DES Y3 3$\times$2pt}}   \\
        \hline
		Flat-$\Lambda$CDM & \mflcdmSNBpt  &  - & - & - & 2194/2212=0.992\\ 
		$\Lambda$CDM & \mlcdmSNBpt & \klcdmSNBpt & - & - & 2194/2211=0.992\\ 
		Flat-$w$CDM & \mfwcdmSNBpt & - & \wfwcdmSNBpt & - & 2188/2211=0.989 \\ 
    	Flat-$w_0w_a$CDM & \mfwaSNBpt & - & \wfwaSNBpt & \wafwaSNBpt & 2191/2210=0.992 \\ 
        \hline
        \multicolumn{6}{l}{\textbf{DES-SN5YR + Planck 2020 + SDSS BAO and DES Y3 3$\times$2pt}}   \\
        \hline
		Flat-$\Lambda$CDM & \mflcdmSNPBpt  &  - & - & - & 2791/2828=0.987\\
		 $\Lambda$CDM & \mlcdmSNPBpt & \klcdmSNPBpt & - & - & 2825/2827=0.999\\  %**todo values and chi2
		Flat-$w$CDM & \mfwcdmSNPBpt & - & \wfwcdmSNPBpt & - & 2785/2827=0.985 \\
    	Flat-$w_0w_a$CDM & \mfwaSNPBpt & - & \wfwaSNPBpt & \wafwaSNPBpt & 2782/2826=0.984 \\ 
        \hline
    \end{tabular}
\end{table*}

\subsection{Goodness of fit and tension}\label{sec:tensions}
\subsubsection{$\chi^2$ per degree of freedom}\label{sec:chi2}
To assess whether our best fits are good fits we calculate the $\chi^2$ per degree of freedom for all our dataset and model combinations; see the last column of Table~\ref{tab:cosmo_results}.  
The $\chi^2$ we use for this test is the maximum likelihood of the entire parameter space, not the marginalized best fit for each parameter. 

The number of degrees of freedom is the number of data points minus the number of parameters that are common to all datasets (i.e., the cosmological parameters of interest). The number of data points added by the CMB, BAO, and 3$\times$2pt is respectively 615, 8, and 471. Due to our treatment of contamination (by inflating the uncertainties of SNe with a low $P_{\rm Ia}$), we approximate the \textit{effective} number of data points in the DES-SN5YR sample by $\sum \PBIa=1735$ (rather than the total number of data points, \numhubble).

Ideally, a good fit should have $\chi^2/$d.o.f.$\sim1.0$. The slightly low $\chi^2/$d.o.f.\ for the DES-SN5YR data arises because $\sum \PBIa$ only approximates the number of degrees of freedom, and the same behaviour is also seen in simulations. 

\subsubsection{Suspiciousness}\label{sec:suspicious}
Suspiciousness, $S$, \citep{handley19} is closely related to the Bayes ratio, $R$,\footnote{Suspiciousness, $S$, is related to the Bayes ratio $R$ and Bayesian information $I$ and is defined as $\ln S= \ln R - \ln I$.} and can be used to assess whether different datasets are consistent. However, while the Bayes ratio has been shown to be prior-dependent \citep{handley19}, with wider prior widths boosting the confidence, Suspiciousness is prior independent. Therefore, Suspiciousness is ideal for cases such as ours where we have chosen deliberately wide and uninformative priors \citep[][Sec.~4.2]{lemos21}.  
\citet{2008ConPh..49...71T} suggests $\ln S < -5$ is ``strong'' tension, $-5 < \ln~S < -2.5$ is ``moderate'' tension, and $\ln S > -2.5$ indicates the datasets are in agreement. 
 
We determine $\ln S$ using the \texttt{ANESTHETIC} software \citep{anesthetic}, which produces an ensemble of realizations used to estimate sample variance. Results are quoted using the mean of the ensemble, with the error bars reflecting the standard deviation.

\begin{figure}\centering
    \includegraphics[width=0.47\textwidth]{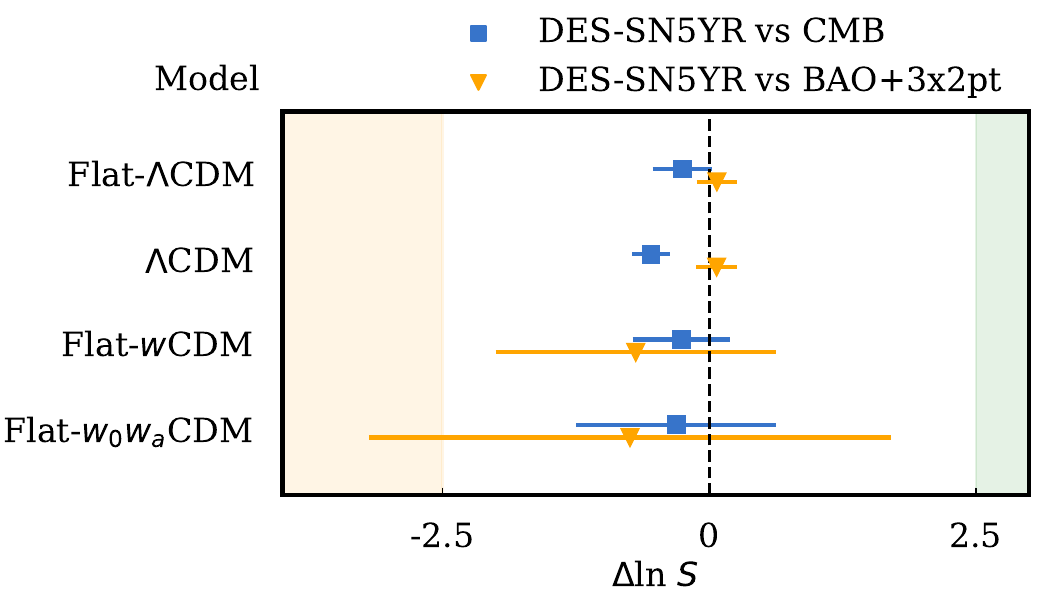}
    \caption{Measurements of Suspiciousness $(\Delta\ln(S))$ between the DES-SN5YR and Planck 2020 datasets for the four models constrained in this paper. Further left indicates higher tension where the shaded regions reflect ``moderate'' (yellow) evidence of tension according to  \citet{2008ConPh..49...71T}. The values and uncertainties represent the mean and standard deviation of realizations estimating sample variance using the \texttt{ANESTHETIC} software. }
    \label{fig:tension} 
\end{figure}

In Fig.~\ref{fig:tension} we plot the Suspiciousness values for the DES-SN5YR data \textit{vs} Planck 2020 and \textit{vs} \BAOpt\ data.  We find no indication of tension using any of
the four models investigated in this paper.

\begin{figure}
    \centering
    \includegraphics[width=0.47\textwidth]{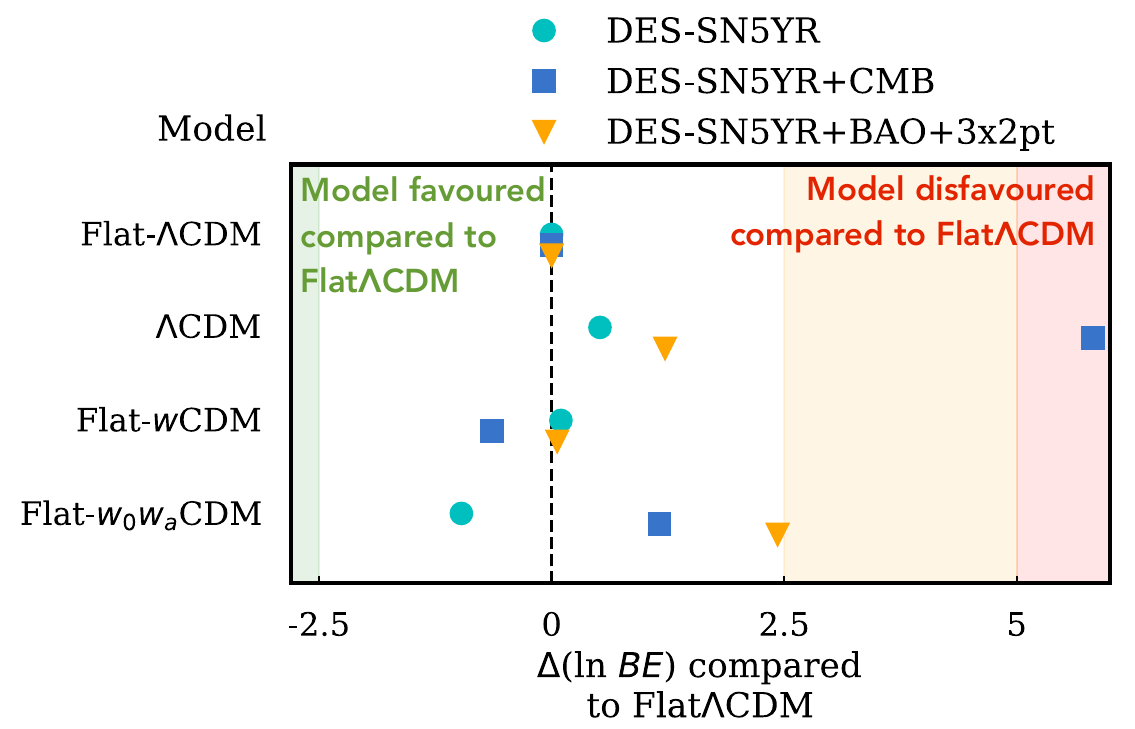}
    \caption{Bayesian Evidence difference relative to Flat-$\Lambda$CDM (\DeltaBE). We present the results for the four different models tested in this analysis and for the three combination of datasets used (DES only in cyan, DES+Planck in blue, DES+\BAOpt\ in orange). An increase (decrease) in \DeltaBE\ indicates that a model is disfavoured (favoured) compared to Flat-$\Lambda$CDM.}
    \label{fig:DeltaBE}
\end{figure}

\subsection{Model Selection}\label{sec:comparison}
Finally, we use Bayesian Evidence to test whether the extra parameters in the more complex models we test are warranted, given the data.  
In Fig.~\ref{fig:DeltaBE}, we present the difference in the logarithm of the Bayesian Evidence, \DeltaBE, relative to Flat-$\Lambda$CDM for the four different models tested in this analysis and for the three combinations of datasets used in Fig.~\ref{fig:DeltaBE}.

To evaluate the strength of evidence
when comparing Flat-$\Lambda$CDM with more complex models, we again use Jeffreys' scale. This empirical scale suggests that \DeltaBE$>2.5$ (and $<-2.5$) is moderate evidence against (in support of) the more complex model, whereas \DeltaBE$>5$ (and $<-5$) is strong evidence against (in support of) the more complex model \citep[for a review of model selection in cosmology see][]{2008ConPh..49...71T}. We note that none of the datasets considered in this analysis strongly favours cosmological models beyond Flat-$\Lambda$CDM. The priors that we choose for model comparison are $w \in (-1.5,-0.5)$, $w_a \in (-10,10)$ and $\ok \in (-0.5,0.5)$. We consider these priors (which determine the penalty for more complex models) to be reasonable in terms of general considerations, such as avoiding universes that are younger than generally accepted stellar ages (see Section \ref{sec:uniage}). Although our chains have been run on uninformative priors, the Bayesian Evidence from those chains may be adjusted for these harmonized priors as described in Appendix \ref{sec:adjustbe}.

%We also find the results to be consistent with the Akaike Information Criteria, another commonly used model comparator. 

\newpage
\section{Discussion}\label{sec:discussion}

\subsection{The big questions}

\subsubsection{Is the expansion of the Universe accelerating?}

Twenty five years ago \citet{riess98} found 99.5\%--99.9\% ($2.8\sigma$ to $3.9\sigma$) evidence for an accelerating Universe, by considering the deceleration parameter $q\equiv(a\ddot{a})\dot{a}^{-2}$ and integrating over the likelihood that $q_0<0$.  Importantly they note that since $q_0$ is measured at the present day but the data span a wide range of redshifts, $q_0$ can only be measured within the context of a model, either cosmographic or physically motivated.  They used the $\Lambda$CDM model, in which $q_0=\om/2-\oll$.  

Doing the same with DES-SN5YR data gives 99.99998\% confidence ($5.2\sigma$) % using "Long Term Sigma" table from https://www.six-sigma-material.com/Tables.html
that $q_0<0$ in $\Lambda$CDM, or a  $2\times10^{-7}$ chance that the expansion of the Universe is {\em not} accelerating. As noted in Section \ref{sec:flatwdcm}, our confidence is even higher that the universe \textit{was} accelerating at $z \sim 0.2$. When we further assume flatness, the confidence in an accelerating Universe is overwhelming (no measurable likelihood for a decelerating Universe) %$>20\sigma$ 
and we find $q_0=$\qFlatcosmographic.  For more fits of $q_0$ using a cosmographic approach see \citet{camilleri24}.

\subsubsection{Is dark energy a cosmological constant?}
\label{sec:cosmoconst}
\label{sec:des_lowz}

As seen in Sec~\ref{sec:constraints}, a cosmological constant is a good fit to our data, but not the best fit.  Our best fit equation of state parameter is slightly (more than $1\sigma$) higher than the cosmological constant value of $w=-1$ (both for SNe alone and in combination with Planck or \BAOpt). 
Our result agrees with the recent result from the UNION3 compilation analyzed with the UNITY framework \citep{rubin23} (which appeared while this paper was under internal review). 
%In contrast to UNION3, which aimed to combine the largest number of supernovae by calibrating them to a common system, we chose to minimise systematics by combining only the best measured and well-calibrated SNe.
The Pantheon$+$ result \citep{brout22_pantheon} is within $1\sigma$ of $w=-1$, but also on the high side ($w=-0.90\pm0.14$).
 
 Furthermore, our analysis slightly prefers a time-varying dark energy equation of state parameter when we fit for $w(a)$ such that the equation of state parameter increases with time (again for all data combinations), known as a ``thawing'' model.  Model selection, however, is inconclusive.

The constraints on time-varying $w$ are enabled by the wide redshift range of the DES-SN5YR sample. 
Our analysis as described in \citet{DES5yr_analysis} gives us confidence that systematic uncertainties in this data are below the level of our statistical precision.  Nevertheless, it is important to recognize that (a) the low-$z$ sample is the one for which we have the least systematic control and (b) the very high-redshift SNe are the ones for which bias-corrections are large ($>0.1$ mag) and more uncertain (e.g., accurate estimation of spectroscopic redshift efficiency is more challenging as we go to higher redshifts), and for which the uncertainties on the rest-frame UV part of the SN~Ia spectral energy distribution have more impact on SN distances estimations \citep[see also][]{brout22_pantheon}. 

To test whether our fits are dominated by any particular redshift range 
we ran cosmological fits (a) removing low-$z$ data (i.e., DES SNe alone) and (b) removing high-$z$ data 
(i.e., removing $\sim80$ SNe at $z>0.85$,  
for which we use only two bands; see Fig.~\ref{fig:lightcurves}).  
Most of the cosmological results obtained with the subsamples are consistent with the results found for the full sample. 
However, we found that removing the low-$z$ sample shifts the contours in the Flat-$w$CDM slightly down, which would make the combined fits more consistent with $w=-1$.  The Flat-$w_0w_a$CDM results are stable to sub-sample selection. See Appendix~\ref{sec:lowz} for details.

We showed in \citet{DES5yr_analysis} that systematic uncertainties are sub-dominant to the statistical uncertainties in our sample.  Nevertheless, in the future a new low-redshift sample (see Sec.~\ref{sec:nextgen}) would help alleviate any remaining doubt about calibration and systematics in the existing low-$z$ sample, and an even higher-redshift supernova survey would help alleviate any modelling concerns by minimizing selection effects even at $z\sim1$.

\subsubsection{How old is the Universe?}
\label{sec:uniage}

One of the issues that the discovery of dark energy solved is the age of the Universe ($t_0$) problem -- globular cluster age estimates, in combination with high estimates of $H_0$, were inconsistent with models that were not accelerating \citep{vandenberg1996GCage_review,gratton1997GCage_Hipparcos,chaboyer1998GC_age_Hipparcos}.

Our results, which favor a dark energy equation of state parameter slightly higher than $w=-1$ would imply that the age is slightly {\em younger} than the age found in a Universe where dark energy is a cosmological constant (for the same values of $H_0$ and present dark energy density).  

To calculate the Universe's age, one needs a value of $H_0$ in addition to the best fit cosmological model.  Since we do not constrain $H_0$ in this analysis, we present our measurement of the combination $H_0t_0$.  In other words, we give $t_0$ in units of the Hubble time $t_H\equiv1/H_0$.\footnote{If $H_0=68$~\kmsMpc, $t_H(68)=14.38$~Gyr.  \\ If $H_0=73$~\kmsMpc, $t_H(73)=13.40$~Gyr.} 
 Our best-fit DES-SN5YR result in Flat-$\Lambda$CDM would have an age of $(0.921\pm0.013) t_H$.  This is $\sim3$\% younger than Planck ($t_{\rm age}^{\rm Planck}=(0.950\pm0.007) t_H$), corresponding to an age difference of approximately $-0.4$~Gyr.    
Our best fit Flat-$w_0w_a$CDM model gives an age $(0.86\pm0.02)t_H$, about 9\% younger than the Flat-$\Lambda$CDM Planck result, corresponding to an age difference of approximately $-1.3$~Gyr.  Such a young age is unlikely given the age of the oldest globular clusters \citep{valcin20,cimatti23,ying23}.  
In the future, this information could be used as a prior to limit the feasible range of time-varying dark energy.  

\subsubsection{Does our best fit resolve the Hubble tension?}
As pointed out in \citet[][their Sec.~5.4]{planck18_VI}, the only basic extensions to the base Flat-$\Lambda$CDM model that resolve the $H_0$ tension are those in which the dark energy equation of state is allowed to vary away from $w=-1$.  
In the $w$CDM model a phantom equation of state parameter of $w\sim-1.5$ would help resolve the tension \citep[][their Sec.~5.1]{diValentino21}, and it is clear from Fig.~\ref{fig:FlatwCDM} that CMB alone actually prefers $w<-1$.  
In this model, Planck alone does not constrain $H_0$ very tightly, and they refrain from quoting a value, 
(see Table~5 of \citet{planck18_VI}), 
but lower $w$ correlates with higher $H_0$. 
However, the DES-SN5YR data shows a slight tendency for $w>-1$, essentially ruling out this solution within $w$CDM.  

\subsection{Comparison with DES-SN3YR and Pantheon+}
\label{disc:DESspec}
It is informative to compare the results of the previous DES-SN3YR analysis \citep{DES-SN3YR, brout19_DES3YR} 
with the results of the DES-SN5YR analysis presented in this work. 
The DES-SN3YR analysis included 207 \textit{spectroscopically confirmed} SNe~Ia from DES and 127 low-redshift SNe from CfA and CSP samples (see also Fig.~\ref{fig:histogram}).
A fraction of those events is in common between both analyses (55 from low-$z$ external samples and 146 DES SNe).\footnote{Not all events included in the DES-SN3YR analysis are included in the DES-SN5YR analysis and vice-versa. This is due to the two analyses implementing different sample cuts. For example the $z>0.025$ cut and the requirement for a host-galaxy redshift in DES-SN5YR exclude respectively 44 and 29 low-$z$ SNe that were in the DES-SN3YR sample. DES-SN5YR also uses a new SALT model (which affects the SALT-based cuts), and is restricted to SNe that pass selection cuts across all systematic tests \citep[see Table 4 in][]{DES5yr_analysis}.}

\begin{figure}\centering
    \includegraphics[width=0.99\linewidth]{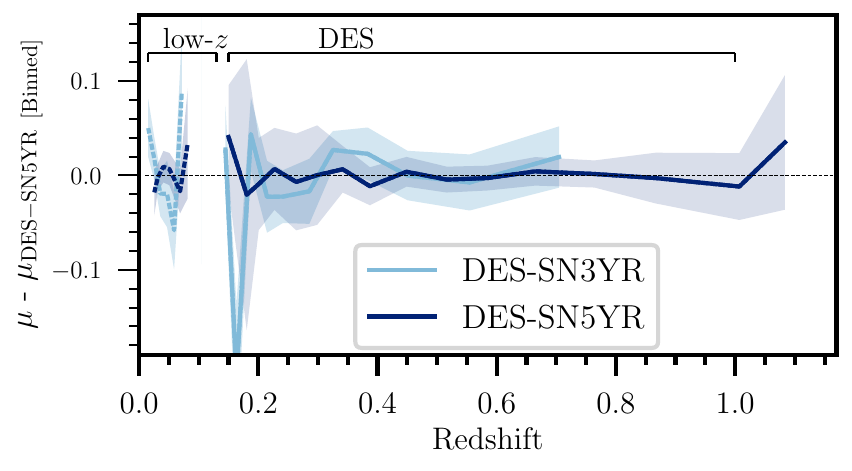}
    \caption{Comparison between Hubble residuals for the DES-SN3YR and DES-SN5YR analyses w.r.t.\ the best fit Flat-$w$CDM for the DES-SN5YR analysis. Hubble residuals are binned in redshift and we present the weighted mean and standard deviation of the mean in each redshift bin. The redshift range covered by the low-$z$ sample is highlighted and shown with thick dotted lines. The two DES samples are consistent with each other. Note the DES-SN3YR analysis only includes spectroscopically confirmed SNe whereas the DES sample in the DES-SN5YR analysis consists entirely of photometrically identified SNe~Ia and extends to higher-$z$.}
    \label{fig:d3d5_comp}
\end{figure}

However, the DES-SN3YR analysis differs from the analysis presented here in many aspects. The SN~Ia intrinsic scatter modelling has been significantly improved \citep[from \lq G10\rq\ and constant $\sigma_{\rm int}$ floor, to the more sophisticated modelling of intrinsic scatter introduced by][]{brout21_BS20,popovic21_dust2dust}, the BBC software has been updated (from BBC \lq 5D\rq\ and a binned approach, to BBC \lq 4D\rq\ and an unbinned approach), the $x_1-M_{\star}$ correlations have been incorporated into simulations \citep[following the work by][]{smith20,popovic21_host}, and the light-curve fitting model has been updated from the SALT2 model to the SALT3 model \citep[see][for a comparison between SALT2 and SALT3 using the DES-SN3YR sample]{taylor23}. Finally, the DES-SN3YR analysis did not require machine-learning classification and the implementation of the BEAMS approach because it is a sample of spectroscopically selected SNe~Ia.
%Therefore, the only quantities that we can directly compare between the two analyses are the
We compare the final SN distances in Fig.~\ref{fig:d3d5_comp} and find consistent results (differences in binned distances are on average 0.02~mag, even in the redshift ranges where contamination is expected to be high). The cosmological results from DES-SN3YR and DES-SN5YR are consistent within uncertainties (when assuming Flat-$\Lambda$CDM, $\om$ are $0.331\pm 0.038$ and \mflcdmSN\ for DES-SN3YR and DES-SN5YR respectively, while when assuming Flat-$w$CDM and including CMB priors, $w$ are $-0.978\pm 0.059$ and \wfwcdmSNP).

\begin{figure}\centering
    \includegraphics[width=0.48\textwidth]{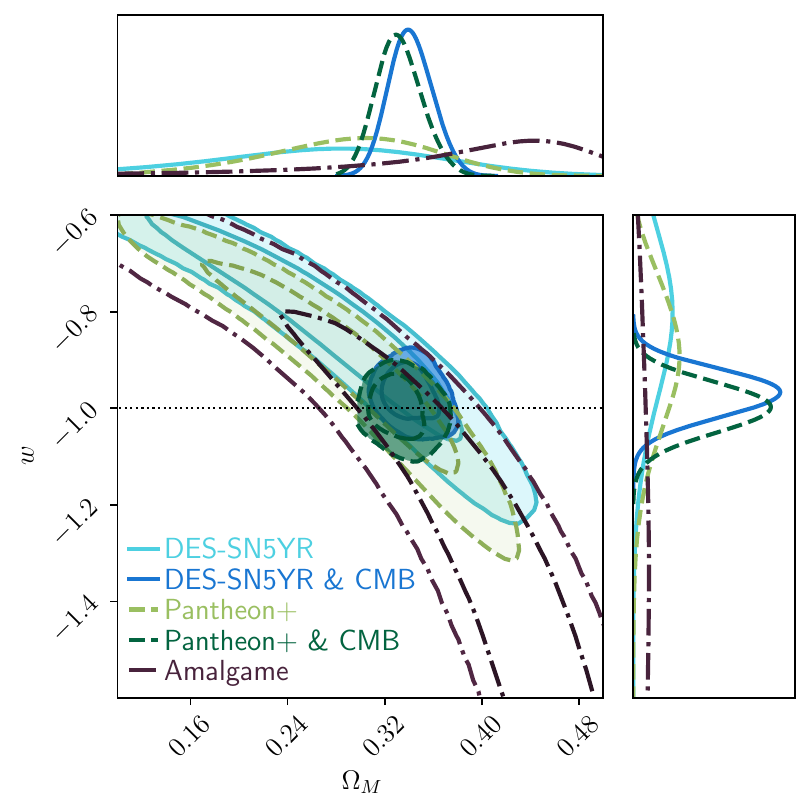}
    \caption{Constraints in Flat-$w$CDM from the DES-SN5YR sample, the Pantheon+ sample (with and without CMB priors), and the Amalgame sample. The constraining power of the  DES-SN5YR and Pantheon+ samples is comparable and consistent, despite Pantheon+ being a spectroscopic SN~Ia sample combining 17 different surveys. The ``Amalgame'' sample includes the SDSS and PS1 photometric SN samples ($>1700$ intermediate-redshift and high-redshift SNe), however it does not include a low-$z$ anchoring sample (hence the larger contours). DES-SN5YR and Pantheon+ are also combined with CMB constraints \citep[for both we use the Planck lite Python implementation presented by][]{prince19}.  The horizontal dotted line marks the equation of state values for a cosmological constant. }
    \label{fig:DES_vs_Ppuls}
\end{figure}

The other main dataset we can compare to is Pantheon+, which contains a significant amount of independent data (all the high-$z$ data).  The DES sample is on average much higher redshift than the Pantheon+ sample (see Fig.~\ref{fig:histogram}), with over a quarter of the DES-SN5YR sample being at high enough redshift ($z\gtrsim0.64$) to probe the likely {\em decelerating}\footnote{The redshift at which the Universe began accelerating in $\Lambda$CDM is $z_{\rm acc} = (2\oll/\om)^{1/3} - 1$.} period of the Universe (compared to 6\% in Pantheon+).  We show a comparison of the contours in Fig.~\ref{fig:DES_vs_Ppuls}.  We find very similar constraining power between Pantheon+ and DES-SN5YR, and the DES-SN5YR value of $w$ is within $1\sigma$ of Pantheon+ \citep{brout22_pantheon}.  These analyses are not fully independent as a fraction of the low-$z$ sample is shared. However, all of the high-$z$ dataset is independent, and DES is a photometric sample while Pantheon+ is fully spectroscopic.  The constraints on $w$ are similar between DES and Pantheon+ as DES high-$z$ has better precision per SN than Pantheon+ and has significantly higher statistical power at $z>0.4$ (see Fig.~\ref{fig:histogram}), but Pantheon+ used $2\times$ more low-redshift SNe (which we do not include in order to be able to better control systematic uncertainties).

\subsection{DES and Next Generation Supernova Samples}\label{sec:nextgen}

This analysis has shown that moving from a spectroscopically confirmed sample as done in \cite{DES-SN3YR} to a photometric sample can increase the sample size of well-measured supernovae significantly (from 207 DES SNe Ia in DES-SN3YR to $>1600$ in DES-5YR), consistent with an analysis of Pan-STARRS SNe in \cite{jones18}.  This improvement arises because photometric classification alleviates the bottleneck of limited spectroscopic resources.  The improvement will increase for future surveys as more candidates are discovered, but the available time for spectroscopy does not increase commensurately.  Importantly, the work of \cite{DES5yr_analysis} shows that systematic uncertainties due to photometric classification are not limiting.  Instead, the ``conventional'' systematics of calibration and modeling the intrinsic scatter remain the most significant challenges.

There is potential for further increase of the statistical power of the DES sample if one moves to using SNe in which a host galaxy spectroscopic redshift was not acquired and instead relies on photometric redshifts of the SNe and the galaxy.  This path was explored by \cite{chen22} for a subset of DES SNe, namely ones that occur in redMaGiC galaxies, and has been explored as well for SuperNova Legacy Survey \citep[SNLS,][]{2022JCAP...10..065R} and the Vera C. Rubin Observatory Legacy Survey of Space and Time (LSST) in \cite{mitra23}.  These analyses show that the use of photo-$z$s do not introduce systematic uncertainties to a scale similar to the statistical uncertainties. This potential is highlighted by the $\approx 2400$ SNe~Ia identified without host galaxy spectroscopic redshift in DES that could be used for this type of analysis \citep{moller24}. %(M\"oller et al. in prep.). 

The DES supernova survey was supported by the 6-year OzDES survey on the Anglo-Australian Telescope \citep[described in][]{lidman20}, which took multi-fibre observations of host galaxies to acquire redshifts of host galaxies of SNe.  The total investment of this program was 100 nights, and for roughly 75\% of the targeted host galaxies a spectroscopic redshift has been secured.  This program was fortuitous as the cameras for OzDES and DECam have a nearly identical field-of-view.  Enormous resources would be needed to reproduce this joint program for LSST, which will find millions of SNe across 18,000 square degrees \citep{ivesic19_lsst,sanchez22}  (compared to the 27 square degrees of DES SNe).   Surveys such as 4MOST will follow-up tens of thousands of these \citep{swann19}, but the full wealth of transient information may benefit from an entirely photometric approach.

As statistical precision continues to improve thanks to the increased number of supernovae, a main theme for systematic analysis is second-order relations between different systematics.  Typically, systematics are treated independently when building the covariance matrix.  We have implemented a method to account for calibration systematics along with light-curve model systematics together, but this is currently the only joint exercise.  This type of work will grow in importance.  For example, while photometric classification does not directly cause a large increase in the error budget, it hinders the ability to constrain the intrinsic scatter model preferred by the data.  Potentially, if LSST and other surveys such as those enabled by the Nancy Grace Roman Space Telescope have enough supernovae \citep{rose21}, the dataset 
can enable a forward modeling approach such as the Approximate Bayesian Computation method introduced in \cite{jennings16} and worked on in Armstrong et al.\ (in prep), which could vary all systematics, nuisance, and cosmological parameters at the same time to compare against the data. 

Furthermore, as discussed in Section~\ref{sec:des_lowz}, modeling of the low-$z$ sample remains a source of systematic uncertainty.  This sample comes from a multitude of surveys, even though we have removed many of the older inhomogeneous sources compared to analyses like Pantheon+.  In the near future, we expect additions from Zwicky Transient Factory \citep{smith24}, Young Supernova Experiment \citep{jones21, aleo23}, and Dark Energy Bedrock All-sky Supernova Survey (DEBASS, PI: Brout) to improve low-$z$ constraints of the SN Hubble Diagram,  given their improved calibration and better understood selection function.\footnote{These upcoming low-$z$ surveys are magnitude-limited rather than targeted, therefore they provide SN samples with a well defined selection function.}  DEBASS will be particularly fruitful as it is a low-redshift sample taken with DECam, 
so a single instrument and calibration catalog will be used for the full sample of DEBASS+DES, 
similar to the single-instrument PS1 sample
in \cite{jones19}.  Using simulations, we estimate that quadrupling the size of our low-$z$ sample (from $\sim200$ to $\sim800$ SNe expected from this next generation of low-$z$ SN surveys) 
could enable a reduction of 
uncertainties on $w$ by $\sim30$ per cent (for a Flat$w$CDM model, using SN data alone).

Lastly, we note that while LSST and Roman may help improve a number of these issues, the first data release is still $>3$ years away.  We encourage work with the DES-SN sample as presented here, combined with other samples.
\citet{popovic23} recently showed the ability to combine separate photometric samples (PS1 and SDSS) into the Amalgame sample (also shown in Fig.~\ref{fig:DES_vs_Ppuls}, and a similar analysis can be done by combining DES with these. It is reasonable to expect that with new low-redshift samples, and combination of high-redshift photometric samples, a sample with $>5000$ likely SNe~Ia can be compiled in the very near future.

\section{Conclusions}\label{sec:conclusions}

The DES Supernova survey stands as a groundbreaking milestone in SN cosmology. With a single survey, we effectively tripled the number of observed SNe~Ia at $z>0.2$ and quintupled the number beyond $z>0.5$.  Here we present the unblinded cosmological results, and in companion papers make public the calibrated light curves and Hubble diagram from the full sample of DES Type Ia supernovae \citep{sanchez24,DES5yr_analysis}. 

After combining the \numdeshd\ DES SNe (of which \numdesia\ have a probability $>0.5$ of being a SN~Ia) with \numlowzia\ existing low-$z$ SNe~Ia, we present final cosmological results for four variants on $\Lambda$CDM cosmology, as summarized in Table~\ref{tab:cosmo_results}.  

The standard Flat-$\Lambda$CDM cosmological model is a good fit to our data. When fitting DES-SN5YR alone and allowing for a time-varying dark energy we do see a slight preference for a dark energy equation of state that becomes greater (closer to zero) with time ($w_a<0$) but this is only at the $\sim2\sigma$ level, and Bayesian Evidence ratios do not strongly prefer the Flat-$w_0w_a$CDM cosmology.  

We compare cosmological results from each of our models to results from the CMB analysis of \citet{planck18_VI}.  There are some differences in the best fit values but in each case we find consistency to within 2$\sigma$ and a Suspiciousness statistic that indicates agreement among the datasets.

Critically, the DES-SN5YR analysis shown here demonstrates that contamination due to SN classification and host-galaxy matching is not a limiting systematic for SN cosmology; this opens the path for a new era of cosmological measurements using SN samples that are not limited by live spectroscopic follow-up of SNe. Instead, our analysis shows the SN community that there are other factors that will be crucial for the success of future SN experiments: 
a high-quality low-redshift sample, a robust UV and NIR extension of light-curve fitting models, 
excellent control of selection effects across the entire redshift range, 
and improvement in our understanding of SN~Ia intrinsic scatter properties and the role played by interstellar dust. 

Future work will conclude the Dark Energy Survey by combining these supernova results with the other three pillars of DES cosmology, namely baryon acoustic oscillations, galaxy clustering, and weak lensing.

%% Also note that the akcnowlodgment environment does not support long amounts of text. If you have a lot of people and institutions to acknowledge, do not use this command. Instead, create a new 
%\section{Softwares}
%Lluis moved this to the dedicated section after the acknowledgement

\section*{Acknowledgments}

{\footnotesize 
We acknowledge the following former collaborators, who have contributed directly to this work --- Ricard Casas, Pete Challis, Michael Childress, Ricardo Covarrubias, Chris D'Andrea, Alex Filippenko, David Finley, John Fisher, Francisco Förster, Daniel Goldstein, Santiago González-Gaitán, Ravi Gupta, Mario Hamuy, Steve Kuhlmann, James Lasker, Marisa March, John Marriner, Eric Morganson, Jennifer Mosher, Elizabeth Swann, Rollin Thomas, and Rachel Wolf.

T.M.D., A.C., R.C., S.H., acknowledge the support of an Australian Research Council Australian Laureate Fellowship (FL180100168) funded by the Australian Government, and A.M. is supported by the ARC Discovery Early Career Researcher Award (DECRA) project number DE230100055.
M.S., H.Q., and J.L are supported by DOE grant DE-FOA-0002424 and NSF grant AST-2108094.
R.K.\ is supported by DOE grant DE-SC0009924. M.V.\ was partly supported by NASA through the NASA Hubble Fellowship grant HST-HF2-51546.001-A awarded by the Space Telescope Science Institute, which is operated by the Association of Universities for Research in Astronomy, Incorporated, under NASA contract NAS5-26555. 
L.K. thanks the UKRI Future Leaders Fellowship for support through the grant MR/T01881X/1.
L.G. acknowledges financial support from the Spanish Ministerio de Ciencia e Innovaci\'on (MCIN), the Agencia Estatal de Investigaci\'on (AEI) 10.13039/501100011033, and the European Social Fund (ESF) ``Investing in your future'' under the 2019 Ram\'on y Cajal program RYC2019-027683-I and the PID2020-115253GA-I00 HOSTFLOWS project, from Centro Superior de Investigaciones Cient\'ificas (CSIC) under the PIE project 20215AT016, and the program Unidad de Excelencia Mar\'ia de Maeztu CEX2020-001058-M, and from the Departament de Recerca i Universitats de la Generalitat de Catalunya through the 2021-SGR-01270 grant. R.J.F. and D.S. were supported in part by NASA grant 14-WPS14-0048. The UCSC team is supported in part by NASA grants NNG16PJ34G and NNG17PX03C issued through the Roman Science Investigation Teams Program; NSF grants AST-1518052 and AST-1815935; NASA through grant No. AR-14296 from the Space Telescope Science Institute, which is operated by AURA, Inc., under NASA contract NAS 5-26555; the Gordon and Betty Moore Foundation; the Heising-Simons Foundation; and fellowships from the Alfred P. Sloan Foundation and the David and Lucile Packard Foundation to R.J.F.
We acknowledge the University of Chicago’s Research Computing Center for their support of this work.

Funding for the DES Projects has been provided by the U.S. Department of Energy, the U.S. National Science Foundation, the Ministry of Science and Education of Spain, 
the Science and Technology Facilities Council of the United Kingdom, the Higher Education Funding Council for England, the National Center for Supercomputing 
Applications at the University of Illinois at Urbana-Champaign, the Kavli Institute of Cosmological Physics at the University of Chicago, 
the Center for Cosmology and Astro-Particle Physics at the Ohio State University,
the Mitchell Institute for Fundamental Physics and Astronomy at Texas A\&M University, Financiadora de Estudos e Projetos, 
Funda{\c c}{\~a}o Carlos Chagas Filho de Amparo {\`a} Pesquisa do Estado do Rio de Janeiro, Conselho Nacional de Desenvolvimento Cient{\'i}fico e Tecnol{\'o}gico and 
the Minist{\'e}rio da Ci{\^e}ncia, Tecnologia e Inova{\c c}{\~a}o, the Deutsche Forschungsgemeinschaft and the Collaborating Institutions in the Dark Energy Survey.

The Collaborating Institutions are Argonne National Laboratory, the University of California at Santa Cruz, the University of Cambridge, Centro de Investigaciones Energ{\'e}ticas, 
Medioambientales y Tecnol{\'o}gicas-Madrid, the University of Chicago, University College London, the DES-Brazil Consortium, the University of Edinburgh, 
the Eidgen{\"o}ssische Technische Hochschule (ETH) Z{\"u}rich, 
Fermi National Accelerator Laboratory, the University of Illinois at Urbana-Champaign, the Institut de Ci{\`e}ncies de l'Espai (IEEC/CSIC), 
the Institut de F{\'i}sica d'Altes Energies, Lawrence Berkeley National Laboratory, the Ludwig-Maximilians Universit{\"a}t M{\"u}nchen and the associated Excellence Cluster Universe, 
the University of Michigan, NSF's NOIRLab, the University of Nottingham, The Ohio State University, the University of Pennsylvania, the University of Portsmouth, 
SLAC National Accelerator Laboratory, Stanford University, the University of Sussex, Texas A\&M University, and the OzDES Membership Consortium.

Based in part on observations at Cerro Tololo Inter-American Observatory at NSF's NOIRLab (NOIRLab Prop. ID 2012B-0001; PI: J. Frieman), which is managed by the Association of Universities for Research in Astronomy (AURA) under a cooperative agreement with the National Science Foundation.
 Based in part on data acquired at the Anglo-Australian Telescope. We acknowledge the traditional custodians of the land on which the AAT stands, the Gamilaraay people, and pay our respects to elders past and present. Parts of this research were supported by the Australian Research Council, through project numbers CE110001020, FL180100168 and DE230100055. Based in part on observations obtained at the international Gemini Observatory, a program of NSF’s NOIRLab, which is managed by the Association of Universities for Research in Astronomy (AURA) under a cooperative agreement with the National Science Foundation on behalf of the Gemini Observatory partnership: the National Science Foundation (United States), National Research Council (Canada), Agencia Nacional de Investigaci\'{o}n y Desarrollo (Chile), Ministerio de Ciencia, Tecnolog\'{i}a e Innovaci\'{o}n (Argentina), Minist\'{e}rio da Ci\^{e}ncia, Tecnologia, Inova\c{c}\~{o}es e Comunica\c{c}\~{o}es (Brazil), and Korea Astronomy and Space Science Institute (Republic of Korea).  This includes data from programs (GN-2015B-Q-10, GN-2016B-LP-10, GN-2017B-LP-10, GS-2013B-Q-45, GS-2015B-Q-7, GS-2016B-LP-10, GS-2016B-Q-41, and GS-2017B-LP-10; PI Foley).  Some of the data presented herein were obtained at Keck Observatory, which is a private 501(c)3 non-profit organization operated as a scientific partnership among the California Institute of Technology, the University of California, and the National Aeronautics and Space Administration (PIs Foley, Kirshner, and Nugent). The Observatory was made possible by the generous financial support of the W.~M.~Keck Foundation.  This paper includes results based on data gathered with the 6.5 meter Magellan Telescopes located at Las Campanas Observatory, Chile (PI Foley), and the Southern African Large Telescope (SALT) (PIs M.~Smith \& E.~Kasai).
The authors wish to recognize and acknowledge the very significant cultural role and reverence that the summit of Maunakea has always had within the Native Hawaiian community. We are most fortunate to have the opportunity to conduct observations from this mountain.

The DES data management system is supported by the National Science Foundation under Grant Numbers AST-1138766 and AST-1536171.
The DES participants from Spanish institutions are partially supported by MICINN under grants ESP2017-89838, PGC2018-094773, PGC2018-102021, SEV-2016-0588, SEV-2016-0597, and MDM-2015-0509, some of which include ERDF funds from the European Union. IFAE is partially funded by the CERCA program of the Generalitat de Catalunya.
Research leading to these results has received funding from the European Research
Council under the European Union's Seventh Framework Program (FP7/2007-2013) including ERC grant agreements 240672, 291329, and 306478.
We  acknowledge support from the Brazilian Instituto Nacional de Ci\^encia
e Tecnologia (INCT) do e-Universo (CNPq grant 465376/2014-2).

This research used resources of the National Energy Research Scientific Computing Center (NERSC), a U.S. Department of Energy Office of Science User Facility located at Lawrence Berkeley National Laboratory, operated under Contract No. DE-AC02-05CH11231 using NERSC award HEP-ERCAP0023923.
This manuscript has been authored by Fermi Research Alliance, LLC under Contract No. DE-AC02-07CH11359 with the U.S. Department of Energy, Office of Science, Office of High Energy Physics.

{\bf Post-publication:} We thank Ósmar Rodríguez for pointing out that the uncertainty on $\Omega_\Lambda$ in Sect.~\ref{sec:lcdm} was overestimated (it was $\pm0.17$ in the published version and has now been corrected to $\pm0.10$).  While preparing \citet{DESBAOSN_2025} we noticed a missing minus sign in the $\ok$ value for $\Lambda$CDM in the DES-SN5YR+Planck case, which has been corrected in this version. All online data and tables had the correct values and are unchanged. 
} %end footnotesize


%\vspace{5mm}
\facilities{CTIO:4m, AAT, Gemini:Gillett (GMOS-N), Gemini:South (GMOS-S), Keck:I (LRIS), Keck:II (DEIMOS), Magellan:Baade (IMACS), Magellan:Clay (LDSS3, MagE), SALT}


\software{
\texttt{numpy} \citep{numpy}, 
\texttt{astropy} \citep{astropy13,astropy18}, 
\texttt{matplotlib} \citep{matplotlib}, 
\texttt{pandas} \citep{pandas}, 
\texttt{scipy} \citep{scipy}, 
\texttt{SNANA} \citep{kessler09}, 
\texttt{Pippin} \citep{hinton20}, 
\texttt{ChainConsumer} \citep{hinton16},
\texttt{Source Extractor} \citep{bertin96}, 
\texttt{MINUIT} \citep{James:1975dr}, SuperNNova \citep{moller20}, SCONE \citep{qu21}.
}



\appendix

\section{Data release and how to use the DES-SN5YR data}
\label{sec:release}

Here we explain where to find the data and software necessary to reproduce our analysis. Many of the codes we use are already public (detailed below). %{\new The primary repository for our key data products is on Zenodo at} \url{http://zenodo.org/***}.  We also mirror the key data and post code and tutorials on Github at \url{https://github.com/des-science/DES-SN5YR}. 
The key data, code, and tutorials are available on Github at \url{https://github.com/des-science/DES-SN5YR}. 

%\subsection{Analysis pipeline inputs and software}
\newcommand{\URLpippin}{\url{https://github.com/dessn/Pippin}}
\newcommand{\URLSNANA}{\url{https://github.com/RickKessler/SNANA}}
\newcommand{\URLSNN}{\url{https://github.com/supernnova/SuperNNova}}
\newcommand{\URLSCONE}{\url{https://github.com/helenqu/scone}}
\newcommand{\URLSMP}{\url{https://smp???}}
\newcommand{\URLDUST}{\url{https://github.com/djbrout/dustdriver}}
\newcommand{\URLSALT}{\url{https://github.com/djones1040/SALTShaker}}
\newcommand{\URLCOSMOSIS}{\url{https://github.com/joezuntz/cosmosis}}
\newcommand{\URLSNDATAROOT}{\url{https://zenodo.org/records/4015325}}

The DES-SN5YR analysis was run using the \textsc{pippin} pipeline framework
\citep{hinton20}\footnote{\URLpippin}
that orchestrated SNANA codes for simulations, light curve
fitting, BBC, and covariance matrix computation 
\citep[\snana,][]{kessler09},\footnote{\URLSNANA} 
and also integrated photometric classification from 
\citet{moller20}\footnote{\URLSNN} and 
\citet{qu21}.\footnote{\URLSCONE}
Additional analyses codes that run outside the main pipeline include
Scene Model Photometry \citep{DES_SMP}, %\footnote{\URLSMP}
fit to measure the SN population of stretch and color
\citep{popovic21_dust2dust},\footnote{\URLDUST}
SALT3 training \citep{kenworthy21},\footnote{\URLSALT} and
CosmoSIS to fit for cosmological parameters \citep{COSMOSIS}.\footnote{\URLCOSMOSIS}

%The entire DES-SN5YR analysis used the SuperNova ANAlysis software 
%\citep[\snana,][]{kessler09},\footnote{\url{https://github.com/RickKessler/SNANA}} 
%is integrated in the \textsc{pippin} pipeline framework
%\citep{hinton20}.\footnote{\url{https://github.com/dessn/Pippin}} Both software packages are open-source and publicly available. 

%{\new On Zenodo and Github we} 
We release the \textsc{pippin} input files necessary to 
%\begin{itemize}
    %\item \textit{ Simulations} \textsc{pippin} input file: to 
    \textit{(i)} generate and fit all the simulations used in the analysis (both the large ``biasCor'' simulations to calculate bias corrections, and the DES-SN5YR-like simulated samples to validate the analysis);
    %\item \textit{ Analysis} \textsc{pippin} input file:
   \textit{(ii)} reproduce the full cosmological analysis, from light-curve fitting to photometric classification, distance estimates and cosmological fitting.
%\end{itemize}
Auxiliary files are also available within the SNANA library.\footnote{\URLSNDATAROOT.}

The various (intermediate and final) \textit{outputs} of our analysis pipeline are also provided. 
This includes \textit{(i)} light-curve fitted parameters, \textit{(ii)} light-curve classification results, \textit{(iii)} the final Hubble diagram and associated uncertainties covariance matrices, and \textit{(iv)} the cosmology chains.

%\section{Marginalising over $H_0$ and $M$}\label{sec:marginalisation}
%The dependency of luminosity distance on $H_0$ and on the rest of the cosmological parameters can be separated into two, $\dl = (c/H_0)\dl^{\prime}$ where $\dl^{\prime}\equiv(1+z_{\rm obs})|\ok|^{-1/2}S_k(\chi)$.
%Thus $\log{\dl}=\log{(c/H_0)}+\log{\dl^\prime}$.  This shows that log of $H_0$ enters as an additive constant in $\mu$ and is thus degenerate with absolute magnitude $M$.  We therefore marginalise over the combination $\mathcal{M}=M+5\log_{10}(c/H_0)$.  To analytically marginalise  one adjusts the $\chi^2$ calculation of Eq.~\ref{eq:chi2} as follows \citep{goliath01},
%\bea
%\tilde{\chi}^2 &=& -2 \ln \left[\int_{-\infty}^{\infty}  \exp\left(-\frac{\chi^2}{2}\right) d\mathcal{M}\right] \\
%&=& \chi^2-\frac{B^2}{C} + \ln\frac{C}{2\pi}, \\
%B &=& \sum_{i=0}^{n} \Delta \mu \mathcal{C}^{-1}, \\
%C &=& \sum_{i,j=0}^{n} \mathcal{C}^{-1}. 
%\eea

\section{Priors}\label{sec:priors}
Table~\ref{tab:priors} lists the prior ranges for our MCMC chains.  The priors related to external data sets align with the priors in the original papers.  We adapted the prior ranges to enclose the majority of the high likelihood region as appropriate for each data set and model combination. Data-set specific priors are listed in the footnote to the table. 

\begin{deluxetable}{cccc}
\tablecolumns{3}
\tablewidth{18pc}
\tablecaption{Priors$^\S$}
\label{tab:priors}
\tablehead{Parameter & \multicolumn{3}{c}{Prior}}
\startdata
\hline \multicolumn{3}{l}{\textbf{Cosmology - baseline}} \\
$h$ & \multicolumn{2}{c}{Flat} & $(0.55,0.91)$ \\
$\Omega_{\mathrm{m}}$ & \multicolumn{2}{c}{Flat} & $(0.1,0.9)$ \\
$10^9 \mathrm{~A}_s$ & \multicolumn{2}{c}{Flat} & $(0.5,5.0)$ \\
$n_{\mathrm{s}}$ & \multicolumn{2}{c}{Flat} & $(0.87,1.07)$ \\
$\Omega_{\mathrm{b}}$ & \multicolumn{2}{c}{Flat} & $(0.03,0.07)$ \\
$\tau$ & \multicolumn{2}{c}{Gaussian} & $(0.067, 0.023)$ \\
$\Omega_\nu$ & \multicolumn{2}{c}{Flat} & $(0.06, 0.6)$ \\
\hline \multicolumn{3}{l}{\textbf{Lens galaxy bias}} \\
$b_i(i \in[1,4])$ & \multicolumn{2}{c}{Flat} & $(0.8,3.0)$ \\
\hline \multicolumn{3}{l}{\textbf{Lens magnification} } \\
$C_1^1$ & Fixed & 0.42 \\
$C_1^2$ & Fixed & 0.30 \\
$C_1^3$ & Fixed & 1.76 \\
$C_1^4$ & Fixed & 1.94 \\
\hline \multicolumn{3}{l}{\textbf{Lens photo-$z$}} \\
$\Delta z_1^1 \times 10^2$ & \multicolumn{2}{c}{Gaussian} & $(-0.9,0.7)$ \\
$\Delta z_1^2 \times 10^2$ & \multicolumn{2}{c}{Gaussian} & $(-3.5,1.1)$ \\
$\Delta z_1^3 \times 10^2$ & \multicolumn{2}{c}{Gaussian} & $(-0.5,0.6)$ \\
$\Delta z_1^4 \times 10^2$ & \multicolumn{2}{c}{Gaussian} & $(-0.7,0.6)$ \\
$\sigma_{2,1}^1$ & \multicolumn{2}{c}{Gaussian} & $(0.98,0.06)$ \\
$\sigma_{z, 1}^{2,1}$ & \multicolumn{2}{c}{Gaussian} & $(1.31,0.09)$ \\
$\sigma_{z, 1}^{31}$ & \multicolumn{2}{c}{Gaussian} & $(0.87,0.05)$ \\
$\sigma_{\varepsilon, 1}^4$ & \multicolumn{2}{c}{Gaussian} & $(0.92,0.05)$ \\
\hline \multicolumn{3}{l}{\textbf{Intrinsic alignment}} \\
$a_i(i \in[1,2])$ & \multicolumn{2}{c}{Flat} & $(-5,5)$ \\
$\alpha_i(i \in[1,2])$ & \multicolumn{2}{c}{Flat} & $(-5,5)$ \\
$b_{\mathrm{TA}}$ & \multicolumn{2}{c}{Flat} & $(0,2)$ \\
$z_0$ & \multicolumn{2}{c}{Fixed} & 0.62 \\
\hline \multicolumn{3}{l}{\textbf{Source photo-$z$}} \\
$\Delta z_s^1 \times 10^2$ & \multicolumn{2}{c}{Gaussian} & $(0.0,1.8)$ \\
$\Delta z_{\mathrm{s}}^2 \times 10^2$ & \multicolumn{2}{c}{Gaussian} & $(0.0,1.5)$ \\
$\Delta z_8^3 \times 10^2$ & \multicolumn{2}{c}{Gaussian} & $(0.0,1.1)$ \\
$\Delta z_8^4 \times 10^2$ & \multicolumn{2}{c}{Gaussian} & $(0.0,1.7)$ \\
\hline \multicolumn{3}{l}{\textbf{Shear calibration}} \\
$m^1 \times 10^2$ & \multicolumn{2}{c}{Gaussian} & $(-0.6,0.9)$ \\
$m^2 \times 10^2$ & \multicolumn{2}{c}{Gaussian} & $(-2.0,0.8)$ \\
$m^3 \times 10^2$ & \multicolumn{2}{c}{Gaussian} & $(-2.4,0.8)$ \\
$m^4 \times 10^2$ & \multicolumn{2}{c}{Gaussian} & $(-3.7,0.8)$ \\
\hline Model & Parameter & \multicolumn{2}{c}{Prior} \\
\hline
\multicolumn{3}{l}{\textbf{Extended models}} \\
$\Lambda$CDM & $\Omega_{K}$ & Flat & $(-0.5, 0.5)$ \\
Flat-$w$CDM & $w$ & Flat & $(-2, 0)$ \\
\multirow{2}{8em}{Flat-$w_0 w_a$CDM} & $w_0$ & Flat & $(-10, 5)$ \\
& $w_a$ & Flat & $(-20, 10)$ \\
\hline
\enddata
\tablenotetext{$\S$}{\scriptsize We also used some specific variations to the above baseline priors. For the $\Lambda$CDM model using the DES-SN5YR only $\Omega_{K}\in (-1.2,2)$, using DES-SN5YR + SDSS BAO and DES Y3 3$\times$2pt $\Omega_{K}\in (-0.8,0.8)$ and using DES-SN5YR + Planck 2020 + SDSS BAO and DES Y3 3$\times$2pt $\Omega_{K}\in (-0.4,0.4)$. For the Flat-$w$CDM model using DES-SN5YR + Planck 2020 $\Omega_{\mathrm{m}}\in (0.1, 1)$ and finally for the Flat-$w_0 w_a$CDM model using DES-SN5YR + SDSS BAO and DES Y3 3$\times$2pt, $w_0 \in -(2, 0)$.}

\end{deluxetable}

\label{sec:adjustbe}
Bayesian Evidence calculations depend on the choice of prior; larger prior ranges used on the same data and likelihoods lead to lower evidences, sometimes referred to as the {\em complex model penalty}. Therefore in model comparison using evidence calculations, we took care to choose consistent prior ranges that do not unduly inflate this penalty. Bayes' Theorem states,
\begin{equation}
    p(M | D) = \frac{p(D | M) p(M)}{p(D)} \propto p(D | M) \;,
\end{equation}
where $D$ is the data and $M$ is the model, and the proportionality to the Bayesian Evidence $p(D | M)$ follows from assuming no prior preference for any model. Writing the model parameters as $\vec{\theta}$ we can then write, 

\begin{equation}
    p(D|M) =  \int p(D,\vec{\theta} | M) d^N \theta 
    = \int p(D| \vec{\theta} , M) p(\vec{\theta}) d^N \theta 
    =  p(\vec{\theta}) \int p(D| \vec{\theta} , M) d^N \theta \;,
\end{equation}
where the last step assumes a constant prior for each of the $N$ parameters $\theta_i$ of model $M$, that fully encompasses the support of the likelihood function (this is true to a very good approximation for the models that are tested here). Making explicit the dependence of the Bayesian Evidence on the model prior by writing $p(D|M) = BE(\vec{\theta})$, the evidence may then be adjusted for a change in prior volume without recomputing the chains as follows : 
\begin{equation}
    \ln{BE(\vec{\theta}_2}) = \ln{BE(\vec{\theta}_1})  + \ln{p(\vec{\theta}_1)} - \ln{p(\vec{\theta}_2)} \;\;,
\end{equation}
where using $(\theta_{i,\rm{min}}, \theta_{i, \rm{max}})$ for the prior range for each parameter, 
\begin{equation}
    p(\vec{\theta}) = \prod_{i =1}^{N} \frac{1}{\theta_{i,\rm{max}} - \theta_{i, \rm{min}}} \;.
\end{equation}

\section{Tests on subsets of our data}\label{sec:lowz}
The large redshift range of the DES-SN5YR sample provides a strong lever arm on the measurement of any time variation of dark energy.  We therefore check for potential peculiarities at the extremes of our redshift range that are driving the fit toward non-cosmological-constant values.  

\begin{figure}\centering
    \includegraphics[width=0.47\linewidth]{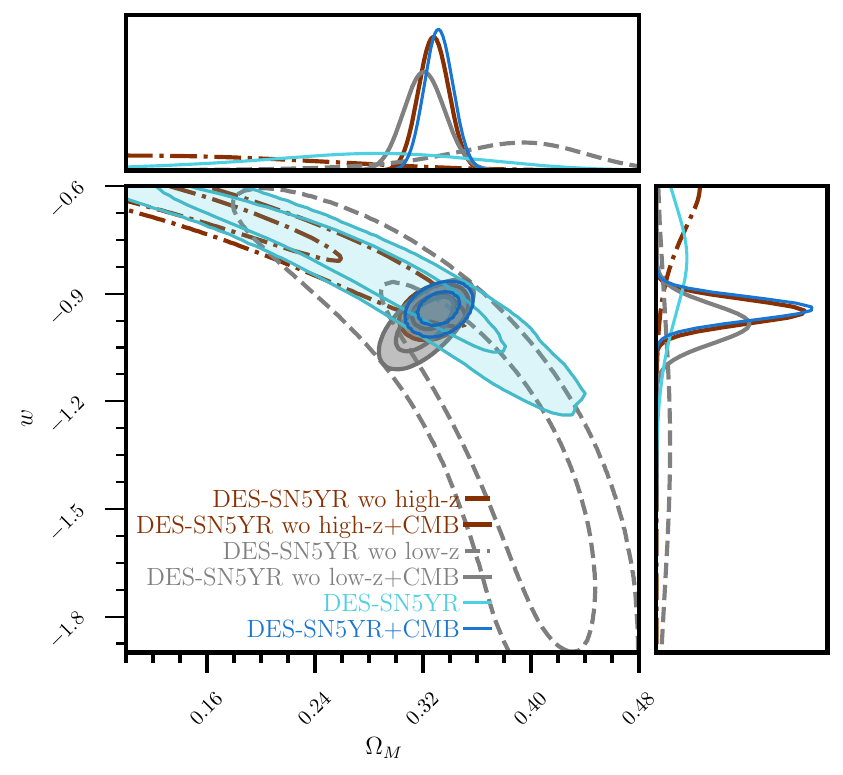}
    \includegraphics[width=0.52\linewidth]{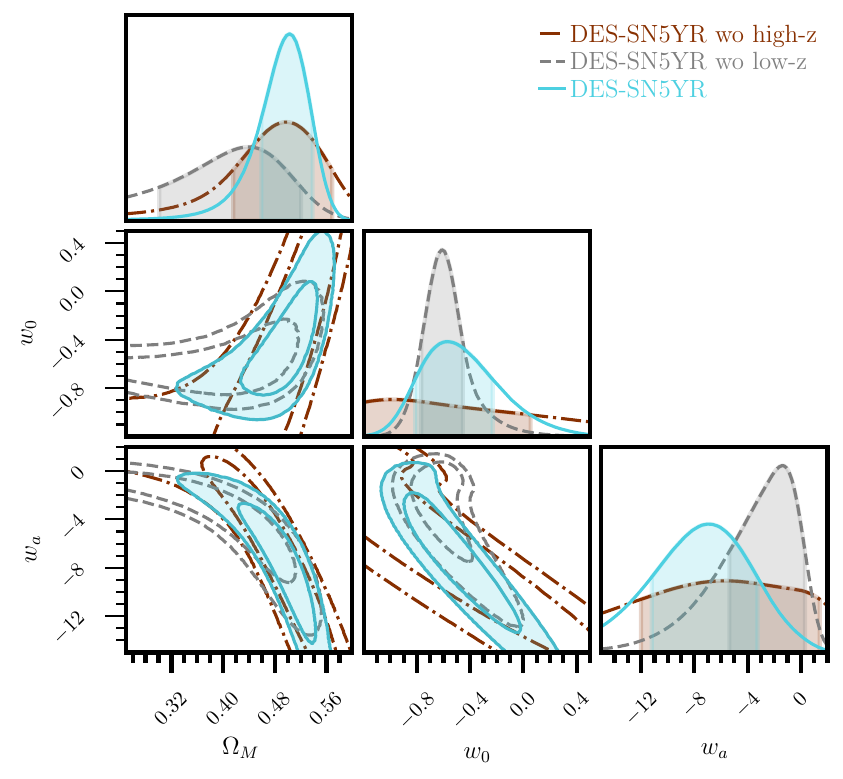}
    \caption{Constraints for the full DES-SN5YR dataset (cyan), when excluding low-$z$ SNe ($z<0.1$, grey dashed line), and when excluding high-$z$ SNe ($z>0.85$, brown dotted-dash line). In Flat-$w$CDM (left) the contours shift primarily along the degeneracy line (and in opposite directions for the low-$z$ and high-$z$ cuts), but also slightly perpendicular to the degeneracy direction.  In combination with the CMB prior this pushes the result closer to $w=-1$ in the no-low-$z$ case. The Flat-$w_0w_a$CDM model (right) best fit sees no significant shifts with sub-sample selection.  
    }
    \label{fig:FlatwCDM_noLowz}
\end{figure}

In Fig.~\ref{fig:FlatwCDM_noLowz} and Table~\ref{tab:cosmo_results_lozhiz}, we show the change to the Flat-$\Lambda$CDM, Flat-$w$CDM and Flat-$w_0w_a$CDM fits using DES-SN alone (no Low-$z$ external samples) and when using the full DES-SN5YR sample but excluding the highest redshift SNe ($z>0.85$, the 5 per cent highest redshift events in our DES SN sample). We show, for example, that in Flat-$\Lambda$CDM excluding the Low-$z$ sample lowers the best fit value to $\om^{{\rm no\, Low-}z}=$\FLCDMomNOLOWZ\ ($\Delta\om=$\FLCDMDomNOLOWZ), which closer agreement with the CMB value of $\om^{\rm Planck}=$\FlcdmPLANCKomegam. Similarly, excluding high redshift SNe lowers the best fit value to $\om^{{\rm no\, High-}z}=$\FLCDMomNOHZ\ ($\Delta\om=$\FLCDMDomNOHZ). However, it is important to quantify the significance of the observed shifts.

The cosmological contours using the full DES-SN5YR sample, the DES-SN5YR sample without Low-$z$, and the DES-SN5YR sample without High-$z$ cannot be \textit{directly compared} as if they were three independent measurements (the three datasets used have large overlaps). 
Therefore, in order to examine the significance of the observed shifts, we generate 100 independent realizations of the DES-SN5YR Hubble diagram applying the Cholesky Decomposition to the full DES-SN5YR data vector of \numhubble\ SNe, and the associated \numhubble$\times$\numhubble\ statistical and systematic covariance matrix. For each independent realization, we fit the cosmological parameters with and without the Low-$z$ and High-$z$ samples and estimate the \textit{standard deviation} ($\sigma$) of the estimated $\Delta\om$ (or $\Delta w$ and/or $\Delta w_0$ and $\Delta w_a$ when fitting for Flat-$w$CDM and Flat-$w_0 w_a$CDM). Using this approach, we measure a $\sigma(\Delta\om)$ of 0.02 and 0.005 when fitting for Flat-$\Lambda$CDM and excluding Low-$z$ and High-$z$ SNe respectively, and we conclude that the $\Delta\om$ observed on the real data are significant at the \FLCDMsigNOLOWZ$\sigma$ and \FLCDMsigNOHZ$\sigma$ respectively. 

\begin{table*}[h]
    \centering
    \caption{Results using DES data alone (excluding Low-$z$ below $z<0.1$) and DES-SN5YR without high-$z$ SNe ($z<0.85$). \textbf{ Shift significance:} the significance of shifts in either $\Omega_M$ (when fitting for Flat$\Lambda$CDM model) or $w$ (when testing  Flat$w$CDM) is estimated from 100 simulations.}
    \label{tab:cosmo_results_lozhiz}
    \begin{tabular}{lcccc}
        \hline
        & $\om$ [$\Delta\om$] &  $w_0$ [$\Delta w_0$] & $w_a$ & Shift significance \\ % & $\Delta$AIC, $\Delta$BIC\\        
        \hline
        \multicolumn{5}{l}{\textbf{DES SNe without Low-$z$}}\\
        \hline
		Flat-$\Lambda$CDM & \FLCDMomNOLOWZ\ [\FLCDMDomNOLOWZ] & - & - & \FLCDMsigNOLOWZ$\sigma$ in $\Omega_M$ \\ 
		Flat-$w$CDM  &  \wCDMomNOLOWZ & \wCDMwNOLOWZ\  [\wCDMDwNOLOWZ] & - &  \wCDMsigNOLOWZ$\sigma$ in $w$ \\ 
		Flat-$w$CDM + \textbf{Planck-like prior} & \CMBwCDMomNOLOWZ & \CMBwCDMwNOLOWZ\ [\CMBwCDMDwNOLOWZ]$^{\dag}$ & - &  \CMBwCDMsigNOLOWZ$\sigma$ in $w$ \\ 
  		Flat-$w_0w_a$CDM  &  \waCDMomNOLOWZ & \waCDMwNOLOWZ\ [\waCDMDwNOLOWZ] & \waCDMwaNOLOWZ\ [\waCDMDwaNOLOWZ] & $<1\sigma$ in $w_0$ and $w_a$ \\ 
    \hline
        \multicolumn{5}{l}{\textbf{DES SNe without High-$z$}}\\
        \hline
            Flat-$\Lambda$CDM & \FLCDMomNOHZ\ [\FLCDMDomNOHZ] & - & - & \FLCDMsigNOHZ$\sigma$ in $\Omega_M$ \\ 
            Flat-$w$CDM  &  \wCDMomNOHZ & \wCDMwNOHZ\ [\wCDMDwNOHZ] & - &  \wCDMsigNOHZ$\sigma$ in $w$ \\ 
            Flat-$w$CDM + \textbf{Planck-like prior} & \CMBwCDMomNOHZ & \CMBwCDMwNOHZ\ [\CMBwCDMDwNOHZ]$^{\dag}$ & - &  \CMBwCDMsigNOHZ $\sigma$ in $w$ \\ 
            Flat-$w_0w_a$CDM  &  \waCDMomNOHZ & \waCDMwNOHZ\ [\waCDMDwNOHZ] & \waCDMwaNOHZ\ [\waCDMDwaNOHZ] & $<1\sigma$ in $w_0$ and $w_a$ \\ 
    \hline
    \end{tabular}
\tablenotetext{$\dag$}{Using the CMB-prior approximation described in the text, we obtain a value of $w=-0.942\pm 0.030$, instead of the value $w=-0.955^{+0.032}_{-0.037}$, presented in Table~\ref{tab:cosmo_results}. For consistency, $\Delta w$ in this table are calculate w.r.t. the $w$-value calculated using the CMB-prior approximation.}
\end{table*}

In Flat-$w$CDM, excluding Low-$z$ gives a best fit $w=$\wCDMwNOLOWZ\  
($\Delta w=$\wCDMDwNOLOWZ) and excluding High-$z$ gives a best fit $w=$\wCDMwNOHZ\  
($\Delta w=$\wCDMDwNOHZ). Using our 100 realizations with systematics, we estimate that the significance of the shifts is \wCDMsigNOLOWZ$\sigma$ and \wCDMsigNOHZ$\sigma$, respectively. 

We perform the same test incorporating a CMB-like prior.
Estimating the best-fit Flat-$w$CDM from our SN subsamples combined with the full CMB likelihood from \citet{planck18_VI} is computationally expensive and practically unfeasible for data and 100 simulations. For this reason, we use an approximation of a CMB-like prior that uses the $R$-parameter \citep[defined, e.g., in][see Eq. 69]{2009ApJS..180..330K} from \citet{planck18_VI}. This CMB-prior approximation is incorporated in the fast minimization cosmological fitting program \texttt{wfit}, available in SNANA.
When combining SNe and the approximated CMB prior and fitting for Flat-$w$CDM, we find that the shifts observed in $w$ are not statistically significant (less than 2$\sigma$).

We make similar tests for Flat-$w_0 w_a$CDM model. The main results are consistent for the different redshift cuts, with the central value varying less than the Flat-$w$CDM case despite (or because of) the extra flexibility of Flat-$w_0w_a$CDM.  

If not statistical fluctuations, the observed shifts in $w$ when removing either low or high-$z$ SNe would be expected if the Flat-$w$CDM model is inadequate and cannot simultaneously fit the both low and high redshift range in our data; but it is also what you expect if there is some kind of systematic error in the low-$z$ or high-$z$ data. Future independent data sets (both supernovae and other measures of expansion such as Baryon Acoustic Oscillations) are essential to determine which is the better explanation. 
The seemingly large values of some of the shifts in cosmological parameters are due to the strong degeneracy in the $w$-$\om$ plane, as seen in Fig.~\ref{fig:FlatwCDM_noLowz}.  Once combined with external data, such as a CMB prior, it is more evident that the shift perpendicular to the degeneracy direction is small (e.g.\ line 3 of Table~\ref{tab:cosmo_results_lozhiz}).

\bibliography{bibliography_new}{}
\bibliographystyle{yahapj_twoauthor_amp} %To allow dual author citation (for DES-5YR DCR paper - J.L.) 


\end{document}